\documentstyle[12pt,oldlfont]{article}
\newcommand{\n}{\hspace*{-2.5mm}}
\newcommand{\simgt}{\rlap{\lower 3.5 pt \hbox{$\mathchar \sim$}} \raise 1pt
 \hbox {$>$}}
\newcommand{\simlt}{\rlap{\lower 3.5 pt \hbox{$\mathchar \sim$}} \raise 1pt
 \hbox {$<$}}

\catcode`@=11
\newcount\@tempcntc
\def\@citex[#1]#2{\if@filesw\immediate\write\@auxout{\string\citation{#2}}\fi
  \@tempcnta\z@\@tempcntb\m@ne\def\@citea{}\@cite{\@for\@citeb:=#2\do
    {\@ifundefined
       {b@\@citeb}{\@citeo\@tempcntb\m@ne\@citea\def\@citea{,}{\bf ?}\@warning
       {Citation `\@citeb' on page \thepage \space undefined}}%
    {\setbox\z@\hbox{\global\@tempcntc0\csname b@\@citeb\endcsname\relax}%
     \ifnum\@tempcntc=\z@ \@citeo\@tempcntb\m@ne
       \@citea\def\@citea{,}\hbox{\csname b@\@citeb\endcsname}%
     \else
      \advance\@tempcntb\@ne
      \ifnum\@tempcntb=\@tempcntc
      \else\advance\@tempcntb\m@ne\@citeo
      \@tempcnta\@tempcntc\@tempcntb\@tempcntc\fi\fi}}\@citeo}{#1}}
\def\@citeo{\ifnum\@tempcnta>\@tempcntb\else\@citea\def\@citea{,}%
  \ifnum\@tempcnta=\@tempcntb\the\@tempcnta\else
   {\advance\@tempcnta\@ne\ifnum\@tempcnta=\@tempcntb \else \def\@citea{--}\fi
    \advance\@tempcnta\m@ne\the\@tempcnta\@citea\the\@tempcntb}\fi\fi}
\catcode`@=12
\begin{document}
\title{\vskip-3cm{\baselineskip14pt
\centerline{\normalsize DESY 95--048\hfill ISSN 0418-9833}
\centerline{\normalsize MPI/PhT/95--20\hfill}
\centerline{\normalsize hep-ph/9503207\hfill}
\centerline{\normalsize March 1995\hfill}}
\vskip1.5cm
Pion and Kaon Production in $e^+e^-$ and $ep$ Collisions at Next-to-Leading
Order}
\author{J. Binnewies$^1$, B.A. Kniehl$^2$, and G. Kramer$^1$\\
$^1$ II. Institut f\"ur Theoretische Physik\thanks{Supported
by Bundesministerium f\"ur Forschung und Technologie, Bonn, Germany,
under Contract 05~6~HH~93P~(5),
and by EEC Program {\it Human Capital and Mobility} through Network
{\it Physics at High Energy Colliders} under Contract
CHRX-CT93-0357 (DG12 COMA).},
Universit\"at Hamburg\\
Luruper Chaussee 149, 22761 Hamburg, Germany\\
$^2$ Max-Planck-Institut f\"ur Physik, Werner-Heisenberg-Institut\\
F\"ohringer Ring 6, 80805 Munich, Germany}
\date{}
\maketitle
\begin{abstract}

We present new sets of fragmentation functions for charged pions and kaons,
both at leading and next-to-leading order.
They are fitted to data on inclusive charged-hadron production in $e^+e^-$
annihilation taken by TPC at PEP ($\sqrt s=29$~GeV) and to similar data by
ALEPH at LEP, who discriminated between events with charm, bottom, and
light-flavour fragmentation in their charged-hadron sample.
In contrast to our previous analysis, where we only distinguished between
valence-quark, sea-quark, and gluon fragmentation, we are now able to treat
all partons independently and to properly incorporate the charm and bottom
thresholds.
Due to the sizeable energy gap between PEP and LEP, we are sensitive to the
scaling violation in the fragmentation process, which allows us to extract a
value for the asymptotic scale parameter of QCD, $\Lambda$.
Recent data on inclusive charged-hadron production in tagged three-jet events
by OPAL and similar data for longitudinal electron polarization by ALEPH
allow us to pin down the gluon fragmentation functions.
Our new fragmentation functions lead to an excellent description of a
multitude of other $e^+e^-$ data on inclusive charged-hadron production,
ranging from $\sqrt s=5.2$~GeV to LEP energy.
In addition, they agree nicely with the transverse-momentum spectra of single
charged hadrons measured by H1 and ZEUS in photoproduction at the $ep$
collider HERA, which represents a nontrivial check of the factorization
theorem of the QCD-improved parton model.
In this comparison, we also find first evidence for the interplay between the
direct- and resolved-photon mechanisms and for the existence of a gluon
density inside the photon.

\end{abstract}
\newpage

\section{Introduction}

The advent of precise data on inclusive hadron production at
$p{\bar p}$, $ep$, and $e^+e^-$ colliders offers the challenging
opportunity to test quantitatively the QCD-improved parton model.
The inclusive cross section is expressed as
a convolution of the parton density functions (PDF), the
partonic cross sections, and the fragmentation functions (FF) of
the quarks and gluons into the outgoing hadrons. The
factorization theorem ensures that the PDF and FF
are universal and that only the hard-scattering partonic cross
sections change when different processes are considered.
While there exist highly sophisticated sets of the PDF for the incoming
protons and photons,
which are needed for the analysis of inclusive particle production in
$p{\bar p}$ and $ep$ reactions, the status of the quark and
gluon FF is much less advanced. The
most direct way to obtain information on the FF
is to analyse the energy spectrum of the hadrons into which
the jets produced by $e^+e^-$ annihilation fragment.

Recently, we performed such an analysis to extract
FF for charged pions and kaons at leading order (LO) and next-to-leading order
(NLO) in QCD \cite{bkk}. These FF were generated through fits to
$e^+e^-$-annihilation data taken at centre-of-mass (CM) energy
$\sqrt s= 29$ GeV by the TPC Collaboration at SLAC \cite{tpc}.
About one year ago, these were the most precise data for
charged-pion and -kaon production.
Our FF also led to rather satisfactory descriptions of other $e^+e^-$
data on charged-particle production at lower (DORIS \cite{dasp,argus}),
similar (PEP \cite{markii}), and higher energies (PETRA \cite{cello,tasso},
TRISTAN \cite{amy}, and LEP \cite{delphi}).

These new parameterizations---in particular, the NLO sets---were tested
against data on single-charged-hadron production
obtained by the H1 Collaboration in $ep$-scattering experiments at HERA
\cite{h1} in Refs.~\cite{h1,lund}
and against similar data obtained in various $p{\bar p}$
experiments in Ref.~\cite{bor}.
The agreement between experimental data and theory turned
out to be very satisfactory. This motivates us to further improve the
FF, especially since
detailed high-statistics data on charged-hadron production at LEP
have become available \cite{aleph,aleph1,opal,lepb}.
For our extended analysis, we shall employ very accurate data from
the ALEPH Collaboration at LEP, who partly
discriminated between pions and kaons \cite{aleph1}.
In the data sample without discrimination of the hadron species,
they distinguished between three cases, namely fragmentation of
(i) $u$, $d$, and $s$ quarks, (ii) $b$ quarks only, and (iii)
all five quark flavours ($u$, $d$, $s$, $c$, and $b$).
In particular, the latter data on the
fragmentation of specific quark flavours enables us to
remove a theoretical assumption which we had to make in our earlier work
\cite{bkk}, namely that the $s$, $c$, and $b$ ($d$, $c$, and $b$) quarks
fragment into charged pions (kaons) in the same way.
As is well known, $b$ quarks are produced considerably more copiously at the
$Z$ resonance than at lower CM energies. Therefore, ALEPH
was able to separate the fragmentation of $b$
quarks into charged hadrons from the fragmentation of all
other quarks.
On the basis of these new ALEPH data on charged-hadron production,
Cowan \cite{aleph} has performed a similar analysis using the
Altarelli-Parisi (AP) equations \cite{ap} for the $Q^2$ evolution and
incorporating data from other experiments at lower energy.
The scope of his analysis is quite different from ours.
He concentrates on charged-hadron data and is chiefly interested in
establishing the scaling violation so as to determine $\alpha_s(M_Z^2)$.
Another significant difference is that he uses a rather high
starting scale, $Q_0=22$~GeV, which lies well above the scale characteristic
for our $ep$ studies.
Similarly to the case of the PDF,
the method of choice for our purposes it to use simple parameterizations
at the low threshold scales.

A problem that requires special attention
is related to the gluon fragmentation into charged
hadrons. Although the gluon does not couple directly to the
electroweak currents, it contributes in higher orders and
mixes with the quarks through the $Q^2$ evolution. This renders it very
difficult to pin down the gluon FF by using just the energy spectrum of
hadrons produced by $e^+e^-$ annihilation.
However, the gluon fragmentation can be probed in $e^+e^-$ annihilation by
exploiting tagged three-jet events or longitudinal polarization. As
in our earlier work \cite{bkk}, we use the information on tagged three-jet
events made available by the OPAL Collaboration \cite{opal} to test our
gluon FF. Recently, in Ref.~\cite{bor},
it was shown that a sufficiently sizeable gluon FF
into charged hadrons is essential to reach good
agreement with the data of single-charged hadron production
in $p{\bar p}$ processes.
Since gluon production is also significant in $ep$ collisions at HERA,
we should be able to further test the strength of the gluon FF extracted from
$e^+e^-$ data by making comparisons with data on inclusive photoproduction of
charged hadrons recently collected by the H1 \cite{h1} and ZEUS Collaborations
\cite{zeus}.

It is the purpose of this work to make use of the recent
data by ALEPH \cite{aleph,aleph1} in addition to the data by TPC \cite{tpc}
and OPAL \cite{opal} to construct new sets of LO and NLO FF
without imposing identities between the FF of different flavours.
(We shall still identify the FF of $u$ and $d$ quarks into charged pions and
those of $u$ and $s$ quarks into charged kaons.)
The new FF sets will be tested against additional $e^+e^-$ data from LEP and
$e^+e^-$ colliders with lower CM energies. In addition, we shall study
single-charged hadron production in $ep$ collisions
under HERA conditions in order to check the consistency of our FF with
the new H1 \cite{h1} and ZEUS data \cite{zeus}, in particular with
respect to the issue of the gluon FF.

This work is organized as follows. In Section~2, we shall
introduce the formalism needed to extract the FF from $e^+e^-$ data on
inclusive single-hadron production at NLO. In Section~3, we shall present
the formulae that describe inclusive single-hadron production
in $ep$ collisions with almost real photons. Section~4 will deal with
the actual analysis and the discussion of our results. In
Section~5, we shall check our results against $e^+e^-$ data
which we did not use in our fits and against the recent H1 and ZEUS data.
Our conclusions will be summarized in Section~6.
As in our earlier work \cite{bkk}, we shall list, in the Appendix, simple
parameterizations of our FF sets.

\section{Formalism for $e^+e^-$ reactions}

The inclusive production of a hadron $h$ by $e^+e^-$ annihilation,
\begin{equation}
\label{process}
e^+e^-\to(\gamma,Z)\to h+X,
\end{equation}
is completely characterized by three observables.
These are the energy fraction of the outgoing hadron
$x=2E_h/\sqrt{s}$,
its angle with the beam axis $\theta$ and the total energy of the system
$\sqrt s$.
The cross section of process~(\ref{process}) exhibits the angular structure
\begin{equation}
\label{sigxtheta}
\frac{d^2\sigma(e^+e^-\to h+X)}{dx\,d{\cos}\theta}=\frac{3}{8}
     (1+{\cos}^2\theta)\frac{d\sigma^T}{dx}
     +\frac{3}{4}{\sin}^2\theta\frac{d\sigma^L}{dx}
     +\frac{3}{4}{\cos}\theta\frac{d\sigma^A}{dx},
\end{equation}
where the superscripts $T$ and $L$ denote the contributions due to
transverse and longitudinal polarizations, respectively, and the asymmetric
term, labelled $A$, accounts for the interference of the photon with the $Z$
boson.
Usually, experiments do not determine $\theta$ distributions.
Thus we shall integrate over $\theta$.
This eliminates the asymmetric term in Eq.~(\ref{sigxtheta}).

The partonic subprocesses may be exactly treated in perturbative QCD.
However, this is not yet possible for the process of hadronization.
As explained in the Introduction, the latter is described in terms
of phenomenological FF, which must be extracted
from experiment.
In the language of the QCD-improved parton model, the $x$ distribution of
process~(\ref{process}) emerges from the $x$ distribution,
$(d\sigma_a/dx)(x,\mu^2,Q^2)$, of $e^+e^-\to a+X$ through
convolution with $D_a^h(x,Q^2)$,
\begin{equation}
\label{sigx}
\frac{1}{\sigma_{tot}}\,\frac{d\sigma(e^+e^-\to h+X)}{dx}=\sum_a
\int\limits_x^1\frac{dz}{z}\,D_a^{h}(z,M_f^2)\,\frac{1}{\sigma_{tot}}\,
\frac{d\sigma_a}{dy}\left(\frac{x}{z},\mu^2,M_f^2\right).
\end{equation}
Here, the sum extents over all active partons ($a=u,d,s,c,b,g$),
$\mu$ is the renormalization scale of the partonic subprocess,
and $M_f$ is the so-called fragmentation scale.
At NLO, $M_f$ defines the point where the divergence associated with collinear
radiation off parton $a$ is to be subtracted.
Throughout this paper, we neglect finite-quark-mass effects in the matrix
elements.
In order to be able to compare our calculations with experimental data,
we normalize the cross section to
\begin{equation}
\label{sigtot}
\sigma_{tot}=N_c\sum_{i=1}^{N_f}e_{q_i}^2
\sigma_0 ,
\end{equation}
where $\sigma_0=(4\pi\alpha^2/3s)$ is the total cross
section of $e^+e^-\to\mu^+\mu^-$ for massless leptons, $N_c=3$,
$e_{q_i}$ is the electroweak charge of quark $q_i$ in units of the positron
charge, and $N_f$ is the number of active flavors.
In Eq.~(\ref{sigtot}), we take into account the running of the fine-structure
constant, $\alpha$, and the enhancement of the contributions due to down-type
quarks at the $Z$ pole.
Since the absolute normalization is determined experimentally only to the 4\%
level, we neglect corrections to $\sigma_{\rm tot}$ in higher orders of
$\alpha_s$.

To NLO in the $\overline{\rm MS}$ scheme, the cross sections of the
relevant subprocesses are given by
\begin{eqnarray}
\label{xsection}
\frac{1}{\sigma_{tot}}\,
\frac{d\sigma_{q_i}}{dy}(y,\mu^2,M_f^2)&\n=\n&
e_{q_i}^2N_c\frac{\sigma_0}{\sigma_{tot}}
\left\{\delta (1-y)+\frac{\alpha_s(\mu^2)}
{2\pi}\left[P_{qq}^{(0,T)}(y)\ln{\frac{s}{M_f^2}}
+K_q^T(y)+K_q^L(y)\right]\right\},\nonumber\\
\frac{1}{\sigma_{tot}}\,\frac{d\sigma_g}{dy}(y,\mu^2,M_f^2)&\n=\n&
2\frac{\alpha_s(\mu^2)}{2\pi}\left[P_{qg}^{(0,T)}(y)\ln{\frac{s}{M_f^2}}
+K_g^T(y)+K_g^L(y)\right].
\end{eqnarray}
$P_{ab}^{(0,T)}(y)$ is the LO term of the transverse $a\to b$
splitting function,
\begin{equation}
\label{split}
P_{ab}^T(y,\alpha_s(Q^2))=P_{ab}^{(0,T)}(y)+\frac{\alpha_s(Q^2)}{2\pi}
P_{ab}^{(1,T)}(y)+\cdots.
\end{equation}
The $K$ functions have been presented in Ref.~\cite{alt}.
The NLO formula of $\alpha_s(\mu^2)$ may be found, {\it e.g.}, in
Ref.~\cite{pdg}.
The scales $\mu$ and $M_f$ are usually identified with the only intrinsic
energy scale, $\sqrt{s}$. For this choice of scales, the terms in
Eq.~(\ref{xsection}) involving $\ln(s/M_f^2)$ are suppressed,
so that the NLO corrections are expressed just in terms of the $K$ functions.
At LO, only the transversely polarized photon contributes
to the cross section.
The longitudinal cross sections of the subprocesses are thus given by
\begin{eqnarray}
\label{sigl}
\frac{1}{\sigma_{tot}}\,
\frac{d\sigma_{q_i}^L}{dy}(y,\mu^2,M_f^2)&\n=\n&
\frac{\alpha_s(\mu^2)}{2\pi}\,C_F\,
e_{q_i}^2N_c\frac{\sigma_0}{\sigma_{tot}}
,\nonumber\\
\frac{1}{\sigma_{tot}}\,\frac{d\sigma_g^L}{dy}(y,\mu^2,M_f^2)&\n=\n&
\frac{\alpha_s(\mu^2)}{2\pi}\,C_F\,\frac{4(1-y)}{y}.
\end{eqnarray}

Having defined the partonic subprocesses, we now turn to the FF.
Their $x$ distributions are not yet calculable in the framework of
perturbative QCD.
However, once we know them at some scale $Q_0^2$,
the $Q^2$ evolution is determined by the AP equations~\cite{ap}.
Our task is thus to construct a model for the $x$ distributions at a
starting scale $Q_0^2$, which, after evolution, fits the data at scale $Q^2$.
The AP equations read
\begin{equation}
\frac{d}{d\ln{Q^2}}D_a^h(x,Q^2)
=\frac{\alpha_s(Q^2)}{2\pi}\sum_{b}\int_x^1{dy\over y}
P_{ba}^T(y,\alpha_s(Q^2))D_{b}^h\left({x\over y},Q^2\right).
\end{equation}
That is, we have to solve a system of integro-differential equations.
This may be achieved with the help of the Mellin-transform technique
\cite{fur}.
The essential property of this transformation is that it renders
convolutions to products.
In fact, denoting the moments of $P_{ab}^T$ and $D_a^h$ by
$A_{ab}$ and $M_a$, respectively, we have
\begin{eqnarray}
\label{glapn}
\frac{d}{d\ln{Q^2}}M_i^+(n,Q^2)&\n=\n&\frac{\alpha_s(Q^2)}{2\pi}
A_{NS}(n,\alpha_s(Q^2))M_i^-(n,Q^2),\nonumber\\
\frac{d}{d\ln{Q^2}}M_{\Sigma}(n,Q^2)&\n=\n& \frac{\alpha_s(Q^2)}{2\pi}
\left[A_{qq}(n,\alpha_s(Q^2))M_{\Sigma}(n,Q^2)
+A_{gq}(n,\alpha_s(Q^2))M_G(n,Q^2)\right],\nonumber\\
\frac{d}{d\ln{Q^2}}M_G(n,Q^2)&\n=\n& \frac{\alpha_s(Q^2)}{2\pi}
\left[A_{qg}(n,\alpha_s(Q^2))M_{\Sigma}(n,Q^2)
+A_{gg}(n,\alpha_s(Q^2))M_G(n,Q^2)\right],\qquad
\end{eqnarray}
where $A_{NS}$ refers to the usual non-singlet combination of quark and
antiquark splitting functions \cite{fur},
the LO expressions for $A_{ab}$ may be found in
Refs.~\cite{fur,flo} and the NLO ones in Ref.~\cite{grv}, and
\begin{eqnarray}
M_i^+(n,Q^2)&\n=\n&\frac{1}{2}\left[M_{q_i}(n,Q^2)
        +M_{\bar{q}_i}(n,Q^2)\right]
        -\frac{1}{2N_f}M_{\Sigma}(n,Q^2),\nonumber\\
M_{\Sigma}(n,Q^2)&\n=\n&\sum_{i=1}^{N_f}
        \left[M_{q_i}(n,Q^2)
        +M_{\bar{q}_i}(n,Q^2)\right],\nonumber\\
M_G(n,Q^2)&\n=\n& M_g(n,Q^2).
\end{eqnarray}
We do not include combinations $M_i^-=M_{q_i}-M_{\bar{q}_i}$,
which are asymmetric under baryon-number symmetry,
since we do not distinguish between quarks and antiquarks.
Note that the non-singlet terms, $M_i^+$, decouple from the gluon.
The solutions of Eq.~(\ref{glapn}) factorize,
\begin{equation}
\label{fragevn}
M(n,Q_2^2)=E(Q_1^2,Q_2^2,N_f)\,M(n,Q_1^2).
\end{equation}
The relevant evolution operators, $E(Q_1^2,Q_2^2)$, have been
calculated in Ref.~\cite{fur}.
For instance, in the non-singlet case, the NLO operator reads
\begin{equation}
\label{nsop}
E_{NS}^{(0)}(Q_1^2,Q_2^2,N_f)=\left[\frac{\alpha_s(Q_1^2)}{\alpha_s(Q_2^2)}
\right]^{\frac{2}{\beta_0}A_{NS}^{(0)}}\left[1+
\frac{\alpha_s(Q_2^2)-\alpha_s(Q_1^2)}{2\pi}\left(
\frac{\beta_1}{\beta_0^2}A_{NS}^{(0)}-\frac{2}{\beta_0}A_{NS}^{(1)}
\right)\right],
\end{equation}
where $\beta_0=(33-2N_f)/3$ and $\beta_1=(153-19N_f)/3$. The
corresponding equation for the coupled system exhibits a similar
structure and is omitted here for ease of presentation.
The subtraction in Eq.~(\ref{nsop}) has the effect that, at $Q^2=Q_0^2$, the
FF are equal to the ansatz also at NLO. In this respect, we differ
from the approach of Ref.~\cite{bkk}.
We evolve the FF in three steps. We start with three quark flavours
at the scale $Q_0$ and evolve up to the charm threshold using the
evolution operators $E(Q_0^2,4m_c^2,3)$. There the charm FF is added to
the set of FF, which, in the second step, are evolved to the bottom threshold
employing $E(4m_c^2,4m_b^2,4)$.
In the final step, the FF of all five quark flavours are evolved to the
considered CM energy $Q^2$ with the help of $E(4m_b^2,Q^2,5)$.

Strictly speaking, our QCD formalism is appropriate only for the
strongly produced pions and kaons.
However, a minor fraction of the observed events are due to pions and
kaons that are produced through weak decays of primary $D$ and $B$ mesons.
In this case, it is unclear whether the usual $Q^2$ evolution may be
applied.
In our analysis, we assumed that this may be done.
A more careful study would require that the experiments discriminate
strongly and weakly produced pions and kaons, which was not the case
for the data used here.

\section{Formalism for $ep$ reactions}

In this section, we shall outline the formalism pertinent to inclusive
photoproduction of hadrons with $ep$ colliders, and in particular with HERA.
According to present HERA conditions, $E_e=26.7$~GeV electrons collide with
$E_p=820$~GeV protons in the laboratory frame, so that $\sqrt s=296$~GeV is
available in the CM frame.
Among the experimentalists it has become customary to take the rapidity,
$y_{\rm lab}$, of hadrons travelling in the proton direction to be positive.
The CM rapidity, $y_{\rm CM}$, is related to $y_{\rm lab}$ by
\begin{equation}
y_{\rm CM}=y_{\rm lab}-{1\over2}\ln{E_p\over E_e}.
\end{equation}
In photoproduction, the electron beam acts like a source of quasi-real
photons, so that HERA is effectively operated as a $\gamma p$ collider.
The appropriate events may be discriminated from deep-inelastic-scattering
events by electron tagging or anti-tagging.
The photon flux is well approximated by the Weizs\"acker-Williams formula
\cite{ros},
\begin{equation}
\label{wwa}
f_{\gamma/e}(z)={\alpha\over2\pi}\left[{1+(1-z)^2\over z}
\ln{Q_{\rm max}^2\over Q_{\rm min}^2}
+2m_e^2z\left({1\over Q_{\rm max}^2}-{1\over Q_{\rm min}^2}\right)\right],
\end{equation}
where $z=E_\gamma/E_e$, $Q_{\rm min}^2=m_e^2z^2/(1-z)$, and
$Q_{\rm max}^2=0.01$~GeV$^2$ (0.02~GeV$^2$) for tagged events at H1 (ZEUS).
The cross section of $ep\to h+X$ emerges from the one of $\gamma p\to h+X$
by convolution with $f_{\gamma/e}(z)$.
By kinematics, $z_{\rm min}\le z\le1$,
where $z_{\rm min}=p_T\exp(-y_{\rm CM})/[\sqrt s-p_T\exp(y_{\rm CM})]$.
In our analysis, we shall impose $0.3<z<0.7$ and $0.318<z<0.431$
to be in conformity with the H1 and ZEUS event-selection criteria,
respectively.

It is well known that $\gamma p\to h+X$ proceeds via two distinct mechanisms.
The photon can interact either directly with the partons originating from the
proton (direct photoproduction) or via its quark and gluon content (resolved
photoproduction).
Both contributions are formally of the same order in the perturbative
expansion.
Leaving aside the proton PDF, $G_b^p(x,M_p^2)$, and the FF,
$D_c^h(x,M_h^2)$, which represent common factors,
the LO cross sections are of $O(\alpha\alpha_s)$ in both cases.
In the case of the resolved mechanism, this may be understood by observing
that the $ab\to cd$ cross sections, which are of $O(\alpha_s^2)$, get dressed
by photon PDF, $G_a^\gamma(x,M_\gamma^2)$, whose leading terms are of the
form $\alpha\ln(M_\gamma^2/\Lambda^2)\propto\alpha/\alpha_s$, with
$\Lambda$ being the asymptotic scale parameter of QCD.
Here, $a,b,c,d$ denote quarks and gluons.
In fact, the two mechanisms also compete with each other numerically.
Resolved photoproduction dominates at small $p_T$ and positive $y_{\rm lab}$,
while direct photoproduction wins out at large $p_T$ and negative $y_{\rm
lab}$.

The LO calculation suffers from significant theoretical uncertainties
connected with the freedom in the choice of the renormalization scale, $\mu$,
of $\alpha_s(\mu^2)$ and the factorization scales,
$M_\gamma,M_p,M_h$.\footnote{%
The LO cross section of direct photoproduction does not depend on $M_\gamma$.}
In order to obtain reliable predictions, it is indispensable to proceed to NLO.
We shall first consider resolved photoproduction, which is more involved.
Starting out from the well-known LO cross section of $\gamma p\to h+X$,
one needs to include the NLO corrections, $K_{ab\to c}$, to the
hard-scattering cross sections, to substitute the two-loop formula for
$\alpha_s$,
and to endow $G_a^\gamma$, $G_b^p$, and $D_c^h$ with NLO evolution.
This leads to
\begin{eqnarray}
\label{res}
&&E_h{d^3\sigma(\gamma p\to h+X)\over d^3p_h}=
\sum_{a,b,c}\int dx_\gamma dx_p{dx_h\over x_h^2}\,
G_a^\gamma(x_\gamma,M_\gamma^2)G_b^p(x_p,M_p^2)D_c^h(x_h,M_h^2)
\nonumber\\&&\times
{1\over\pi\hat s}\left[{1\over v}\,{d\sigma_{ab\to c}^0\over dv}
(\hat s,v;\mu^2)\delta(1-w)
+{\alpha_s(\mu^2)\over2\pi}K_{ab\to c}(\hat s,v,w;\mu^2,M_\gamma^2,M_p^2,M_h^2)
\right],
\end{eqnarray}
where $d\sigma_{ab\to c}^0/dv$ are the LO hard-scattering cross sections,
$v=1+\hat t/\hat s$, and $w=-\hat u/(\hat s+\hat t)$, with
$\hat s=(p_a+p_b)^2$, $\hat t=(p_a-p_c)^2$, and $\hat u=(p_b-p_c)^2$ being the
Mandelstam variables at the parton level.
The parton momenta are related to the photon, proton, and hadron momenta by
$p_a=x_\gamma p_\gamma$, $p_b=x_p p_p$, and $p_c=p_h/x_h$.
The indices $a,b,c$ run over the gluon and $N_f$ flavours of quarks and
antiquarks.
In this paper, we shall take $N_f=5$.
The $K_{ab\to c}$ functions may be found in Ref.~\cite{ave} for $M_\gamma=M_p$.
This restriction was relaxed in Ref.~\cite{kni}.

The NLO cross section of direct photoproduction emerges from Eq.~(\ref{res})
by substituting $G_a^\gamma(x_\gamma,M_\gamma^2)=\delta(1-x_\gamma)$,
replacing $d\sigma_{ab\to c}/dv$ and $K_{ab\to c}$ by
$d\sigma_{\gamma b\to c}/dv$ and $K_{\gamma b\to c}$, respectively,
and omitting the sum over $a$.
The $K_{\gamma b\to c}$ functions were first derived in Ref.~\cite{aur}
setting $M_\gamma=M_p=M_h$ and taking the spin-average for incoming
photons and gluons to be 1/2.
In Ref.~\cite{kni}, the scales were disentangled and the spin-average
convention was converted to the $\overline{\rm MS}$ scheme, {\it i.e.}, to be
$1/(n-2)$, with $n$ being the dimensionality of space-time.
Analytic expressions for the $K_{\gamma b\to c}$ functions are listed in
Ref.~\cite{gor}.
Quantitative studies of inclusive photoproduction of hadrons at HERA via the
direct- and resolved-photon mechanisms in NLO may be found in
Refs.~\cite{kni,gor,fmb}.

\section{Results}

In general, data of inclusive hadron production at $e^+e^-$ colliders
are most suitable for the extraction of the FF;
in the case of fixed-target, collider, and $ep$ data, the information
on the FF is obscured by theoretical uncertainties
arising from the PDF and the choice of factorization
scales connected with the initial state.
For our analysis, we select the data on charged-pion and -kaon production
taken at energy $\sqrt s=29$~GeV by the TCP Collaboration at SLAC \cite{tpc}
and those collected at $\sqrt s=M_Z$ by the ALEPH Collaboration at LEP
\cite{aleph1}.
In order to constrain the FF of the different quark flavours, we employ
the ALEPH data \cite{aleph} on charged-hadron production for the three cases:
(1) sum over all quark flavours ($u,d,s,c,b$);
(2) sum over $u,d,s$ quarks; and
(3) $b$ quark.
These data combine small statistical errors with fine binning in $x$
and are thus more constraining than data collected by other
$e^+e^-$ experiments in the energy range from 5.2 to 55~GeV.
For this reason, our approach is to fit exclusively to TPC and ALEPH data and
to use the other data for cross checks.
To control the gluon FF, we exploit, as in our previous work \cite{bkk},
the information on the tagged three-jet events by OPAL \cite{opal}
presented by Nason and Webber \cite{opal}, who applied small corrections to
the original data.
However, these data comprise just five points and are not very constraining.
For the fitting procedure, we use the $x$ bins in the interval between
$x_{\rm min}=\max(0.1,2~{\rm GeV}/\sqrt s)$ and 0.8 and integrate the
theoretical functions over the bin widths as in the experimental data.
The restriction at small $x$ is necessary to exclude events in the
non-perturbative region, where mass effects are important and our formalism
is bound to fail.
We parameterize the $x$ dependence of the FF at $Q_0^2$ as
\begin{equation}
D_a^h(x,Q_0^2)=Nx^{\alpha}(1-x)^{\beta}.
\end{equation}
In contrast to our previous work \cite{bkk}, we now only impose the conditions
$D_d^{\pi^++\pi^-}=D_u^{\pi^++\pi^-}$ and $D_s^{K^++K^-}=D_u^{K^++K^-}$.
For all the other FF, including those of the gluon, we keep $N$, $\alpha$,
and $\beta$ as independent fit parameters.
This means that, together with $\Lambda$, which we also keep as a free
parameter, we have a total of 31 independent fit parameters.

The quality of the fit is measured in terms of the average
$\chi^2_{\rm d.o.f.}$ for all selected data points.
Our technical procedure is as follows.
We consider Eq.~(\ref{sigx}) at $\mu=M_f=\sqrt s=29$~GeV and $M_Z$ as a
function of the 31 parameters that determine the $x$ dependence of the FF
at the respective values of $Q_0$,
which we take to be $Q_0=\sqrt2$~GeV for $u,d,s$,
$Q_0=m(\eta_c)=2.9788$~GeV \cite{pdg} for $c$, and
$Q_0=m(\Upsilon)=9.46037$~GeV \cite{pdg} for $b$.
Using a multidimensional minimization algorithm \cite{min},
we search this 31-dimensional parameter space for the point at which
the deviation of the theoretical prediction from the data becomes minimal.
When we fit to charged-hadron data, we must include the contribution due to
protons and antiprotons.
To this end, we introduce a function, $f(x)$, which parameterizes the ratio
of the cross sections of $p+\bar p$ and $\pi^++\pi^-$ production as measured
by ALEPH \cite{aleph1},
\begin{equation}
\label{fx}
f(x)=0.195-1.35\,(x-0.35)^2.
\end{equation}
Thus, the charged-particle production cross section is given by
\begin{equation}
\label{charge}
{d\sigma^{h^++h^-}\over dx}=[1+f(x)]{d\sigma^{\pi^++\pi^-}\over dx}
+{d\sigma^{K^++K^-}\over dx}.
\end{equation}
To test the perturbative stability, we perform fits in LO and NLO,
although we are primarily interested in the NLO results.

Our results are listed below.
For the average of charged pions, we obtain
\begin{eqnarray}
\label{pilo}
D_u^{(\pi,LO)}(x,Q_0^2)&\n=\n&D_d^{(\pi,LO)}(x,Q_0^2)
=1.09\,x^{-0.842}\,(1-x)^{1.43},\nonumber\\
D_s^{(\pi,LO)}(x,Q_0^2)&\n=\n&3.47\,x^{-1.037}\,(1-x)^{3.88},\nonumber\\
D_c^{(\pi,LO)}(x,Q_0^2)&\n=\n&4.52\,x^{-0.898}\,(1-x)^{4.69},\nonumber\\
D_b^{(\pi,LO)}(x,Q_0^2)&\n=\n&3.62\,x^{-1.128}\,(1-x)^{7.12},\nonumber\\
D_g^{(\pi,LO)}(x,Q_0^2)&\n=\n&6.20\,x^{-0.351}\,(1-x)^{3.00}
\end{eqnarray}
in LO and
\begin{eqnarray}
D_u^{(\pi,NLO)}(x,Q_0^2)&\n=\n&D_d^{(\pi,NLO)}(x,Q_0^2)
=1.16\,x^{-0.762}\,(1-x)^{1.47},\nonumber\\
D_s^{(\pi,NLO)}(x,Q_0^2)&\n=\n&3.96\,x^{-0.825}\,(1-x)^{4.38},\nonumber\\
D_c^{(\pi,NLO)}(x,Q_0^2)&\n=\n&3.69\,x^{-0.873}\,(1-x)^{5.06},\nonumber\\
D_b^{(\pi,NLO)}(x,Q_0^2)&\n=\n&4.00\,x^{-1.037}\,(1-x)^{7.75},\nonumber\\
D_g^{(\pi,NLO)}(x,Q_0^2)&\n=\n&5.11\,x^{-0.287}\,(1-x)^{2.61}
\end{eqnarray}
in NLO.
In the case of kaons, we find
\begin{eqnarray}
D_u^{(K,LO)}(x,Q_0^2)&\n=\n&D_s^{(K,LO)}(x,Q_0^2)
=0.38\,x^{-1.234}\,(1-x)^{1.04},\nonumber\\
D_d^{(K,LO)}(x,Q_0^2)&\n=\n&1.11\,x^{-0.942}\,(1-x)^{2.81},\nonumber\\
D_c^{(K,LO)}(x,Q_0^2)&\n=\n&0.55\,x^{-0.764}\,(1-x)^{2.69},\nonumber\\
D_b^{(K,LO)}(x,Q_0^2)&\n=\n&0.72\,x^{-0.792}\,(1-x)^{2.81},\nonumber\\
D_g^{(K,LO)}(x,Q_0^2)&\n=\n&0.43\,x^{-0.374}\,(1-x)^{2.69}
\end{eqnarray}
in LO and
\begin{eqnarray}
\label{kanlo}
D_u^{(K,NLO)}(x,Q_0^2)&\n=\n&D_s^{(K,NLO)}(x,Q_0^2)
=0.30\,x^{-0.978}\,(1-x)^{1.01},\nonumber\\
D_d^{(K,NLO)}(x,Q_0^2)&\n=\n&1.05\,x^{-0.804}\,(1-x)^{2.54},\nonumber\\
D_c^{(K,NLO)}(x,Q_0^2)&\n=\n&0.68\,x^{-0.753}\,(1-x)^{2.59},\nonumber\\
D_b^{(K,NLO)}(x,Q_0^2)&\n=\n&0.58\,x^{-0.843}\,(1-x)^{2.88},\nonumber\\
D_g^{(K,NLO)}(x,Q_0^2)&\n=\n&0.33\,x^{-0.351}\,(1-x)^{0.65}
\end{eqnarray}
in NLO.
Here, it is understood that the $Q_0^2$ values refer to the individual
starting points given above.
The $\chi^2_{\rm d.o.f.}$ values achieved for the different data sets may
be seen from Table~I.

\begin{table} {TABLE~I. CM energies, types of data, numbers of data points
used, and $\chi^2_{\rm d.o.f.}$ values obtained at NLO and LO for the various
data samples discussed in the text.
The data used in the fits are marked by an asterisk.}\\[1ex]
\begin{tabular}{|c|lc|c|c|c|c|} \hline
$\sqrt s$ [GeV] & Data type &&Ref.\ & No.\ of points & $\chi^2_{\rm d.o.f.}$
in NLO & $\chi^2_{\rm d.o.f.}$ in LO \\
\hline
91.2 & $\sigma^h$ (all flavours) &*& \cite{aleph}  & 23 & 0.84 & 0.85 \\
     & $\sigma^h$ ($u,d,s$)      &*& \cite{aleph}  & 23 & 1.13 & 1.41 \\
     & $\sigma^h$ ($b$)          &*& \cite{aleph}  & 23 & 1.06 & 0.84 \\
     & $\sigma^\pi$              &*& \cite{aleph1} & 13 & 1.19 & 1.30 \\
     & $\sigma^K$                &*& \cite{aleph1} & 13 & 1.10 & 1.57 \\
     & $D_g^h$                   &*& \cite{opal}   & 5  & 7.05 & 6.22 \\
     & $\sigma^h$ (longitudinal) && \cite{aleph}  & 10 & 8.65 & --   \\
     & $\sigma^h$ (all flavours) && \cite{delphi} & 22 & 0.95 & 0.82 \\
\hline
55.2 & $\sigma^h$                && \cite{amy}    &  7 & 1.18 & 1.23 \\
\hline
43.7 & $\sigma^\pi$              && \cite{tasso}  &  3 & 0.94 & 0.92 \\
\hline
35.0 & $\sigma^h$                && \cite{cello}  & 22 & 0.64 & 1.24 \\
\hline
34.0 & $\sigma^\pi$              && \cite{tasso}  &  6 & 0.63 & 0.54 \\
     & $\sigma^K$                && \cite{tasso}  &  2 & 0.66 & 0.99 \\
\hline
29.0 & $\sigma^h$                && \cite{markii} & 14 & 1.55 & 2.13 \\
     & $\sigma^\pi$              &*& \cite{tpc}    & 18 & 1.30 & 1.49 \\
     & $\sigma^K$                &*& \cite{tpc}    & 18 & 0.88 & 0.75 \\
\hline
9.98 & $\sigma^\pi$              && \cite{argus}  & 19 & 1.56 & 1.88 \\
     & $\sigma^K$                && \cite{argus}  & 14 & 1.00 & 1.35 \\
\hline
5.20 & $\sigma^\pi$              && \cite{dasp}   &  5 & 0.93 & 0.45 \\
     & $\sigma^K$                && \cite{dasp}   &  3 & 1.49 & 2.31 \\
\hline
\end{tabular}
\end{table}

For the reader's convenience, we list simple parameterizations of the $x$ and
$Q^2$ dependences of these sets in the Appendix.
We believe that such parameterizations are indispensable for practical
purposes, especially at NLO.
However, we should caution the reader that these parameterizations describe
the evolution of the FF only approximately.
Deviations in excess of 10\% may occur for $x<0.1$, in particular for the
gluon.
While this kind of accuracy is fully satisfactory for most applications, it is
insufficient for the comparison with the high-statistics data collected at LEP.
We wish to point out that all $\chi^2_{\rm d.o.f.}$ values presented in this
paper have been computed with the full AP-evolved results, which
have an estimated relative error of less than 0.4\%.

Our FF exhibit good perturbative stability.
In general, there is only little difference between our LO and NLO sets.
The only appreciable difference occurs for $D_g^K$.
Since we included in our fit high-quality data from two very different
energies, namely 29 and 91.2~GeV, we are sensitive to the running of
$\alpha_s$ and are, therefore, able to extract a value of
$\Lambda_{\overline{\rm MS}}^{(5)}$, appropriate to five active quark flavours.
We find $\Lambda_{\overline{\rm MS}}^{(5)}=195$~MeV
($107$~MeV) in NLO (LO),
which corresponds to $\alpha_s(M_Z^2)=0.1154$
($0.1215$), in good
agreement with the value $0.120\pm0.008$ extracted from a global fit to
the $Z$-boson observables measured at LEP \cite{heb}.
We find that the error on $\Lambda_{\overline{\rm MS}}^{(5)}$ due to
faked scaling effects in connection with the distinct flavour composition at
the $Z$ resonance is of the order of $20$~MeV. We are quite insensitive
to other sources of uncertainties in $\Lambda$, such as the one related to
the relative normalization of the data.

In this context, we also wish to mention nonperturbative power corrections
to the fragmentation process. We assume throughout that such
corrections are small and need not be considered when analyzing
the presently available data. The good agreement between our results and the
various measurements nicely supports this assumption. However,
it cannot yet be excluded theoretically that corrections to inclusive timelike
processes (such as $e^+e^-$ annihilation) might be of order $1/Q$, which
would correspond to a sizeable uncertainty in
$\Lambda_{\overline{\rm MS}}^{(5)}$.

Since we have built in the $c\bar c$ and $b\bar b$ thresholds,
we have three different starting scales $Q_0$.
To illustrate the relative size of the FF, we have plotted them in
Fig.~\ref{fig1}a--d as functions of $x$ for $Q=10$~GeV.
We observe that, in the case of the pion sets, the $b$-quark distribution is
rather soft, whereas the $u$-quark distribution is quite hard.

The goodness of our fits to the TPC \cite{tpc} and ALEPH data
\cite{aleph1} on charged-pion and -kaon production may be judged from
Figs.~\ref{fig2} and \ref{fig3}.
The fit to the charged-hadron data by ALEPH is seen in Fig.~\ref{fig4}, where
we show the contributions due to (1) $u,d,s$, (2) $b$, and (3) all flavours.
We also compare the contribution due to gluon fragmentation into charged
hadrons with the data obtained in the three-jet analysis by OPAL \cite{opal}.
The full (dotted) curves correspond to our NLO (LO) analysis.
In the case of pion and charged-hadron production, our fits describe the data
quite well, even for $x<x_{\rm min}$.
Specifically, we find $\chi^2_{\rm d.o.f.}$ values of
1.28 (1.32) at NLO (LO), if we include the gluon information.
Leaving out the gluon data, we obtain 1.06 (1.14).
As may be seen from Table~I, the five gluon data points
contribute an extraordinarily high value of $\chi^2_{\rm d.o.f.}$.

This concludes the description of our fit procedure and the quality of the
fit to the data used.
In the next section, we shall first study to what extend our FF
also account for other data, both below and at the $Z$ resonance,
including the longitudinal cross section at the $Z$ peak, which has recently
been  measured \cite{aleph}.
In addition, we shall confront our prediction for charged-particle production
in $ep$ scattering at HERA with the data by H1 \cite{h1} and ZEUS \cite{zeus}.

\section{Applications}

In order to gain confidence in the validity of our new FF sets, we shall now
test them against data which were not used in our fits.
Specifically, we shall consider data on charged-pion and -kaon production
taken by the DASP \cite{dasp} and ARGUS Collaborations \cite{argus} at DORIS
and the TASSO Collaboration \cite{tasso} at PETRA.
Furthermore, we shall make comparisons with charged-hadron data collected by
MARK~II \cite{markii} at PEP, CELLO \cite{cello} at PETRA, AMY \cite{amy} at
TRISTAN, and DELPHI \cite{delphi} at LEP.
In the latter case, we shall assume that the percentage of the produced
protons and antiprotons relative to the charged pions may be approximated
for all energies by the function $f(x)$ given in Eq.~(\ref{fx}), so that
Eq.~(\ref{charge}) is valid.
For each of these data samples, we list in Table~I the CM energy, the number
of data points with $x_{\rm min}<x<0.8$, and the resulting
$\chi^2_{\rm d.o.f.}$ values in NLO and LO.
We see that all these data are fitted rather well, with $\chi^2_{\rm d.o.f.}$
values of the order of unity.
In fact, these $\chi^2_{\rm d.o.f.}$ values are comparable with those
we obtained for the TPC \cite{tpc} and ALEPH \cite{aleph,aleph1} data, to which
we actually fitted.
This serves as a nontrivial consistency check for our procedure and tells us
that a global fit to all data would not further improve the goodness of our FF.
The NLO and LO values of $\chi^2_{\rm d.o.f.}$ are very similar, with a slight
preference for our NLO set.
The good agreement with the DASP \cite{dasp}, ARGUS \cite{argus}, and
TASSO \cite{tasso} data of charged-pion and -kaon production is also
demonstrated in Figs.~\ref{fig2} and \ref{fig3}, respectively.
The ARGUS data are also very well described for $x<x_{\rm cut}$.
In Fig.~\ref{fig4}, we present the comparisons of our theoretical results with
the MARK~II \cite{markii}, CELLO \cite{cello}, and AMY \cite{amy} measurements
of charged-hadron production.
We see from Figs.~\ref{fig2}--\ref{fig4} that the LO and NLO calculations
agree very well at $\sqrt s=91.2$~GeV, whereas they tend to deviate at smaller
energies, the discrepancy being up to 30\% at $\sqrt s=5.20$~GeV and medium
$x$.

Another method to extract information on the gluon FF is to analyze the
longitudinal cross section $d\sigma^L/dx$ defined in Eq.~(\ref{sigxtheta}).
Similarly the total cross section $d\sigma/dx=d\sigma^T/dx+d\sigma^L/dx$
defined in Eq.~(\ref{sigx}), which we have considered so far,
$d\sigma^L/dx$ may be represented as a convolution of the FF
with the longitudinal cross sections of the individual subprocesses
given in Eq.~(\ref{sigl}),
\begin{equation}
\label{sigmal}
{1\over\sigma_{\rm tot}}\,{d\sigma^L\over dx}=
{\alpha_s\over2\pi}C_F\int_x^1{dz\over z}
\left[
N_c\frac{\sigma_0}{\sigma_{tot}}
\sum_{i=1}^{2N_f}e_{q_i}^2D_{q_i}^h(z,Q^2)
 + 4\left(\frac{z}{x}-1\right)D_g^h(z,Q^2)
\right]
+O(\alpha_s^2).
\end{equation}
Since Eq.~(\ref{sigmal}) is only given to LO in $\alpha_s$, the $\alpha_s$
value here is not expected to be the same as the one in the NLO
relations (\ref{sigx})--(\ref{split}).
Thus, from measurements of $d\sigma^L/dx$ and $d\sigma^T/dx$, the gluon FF can
only be extracted to LO.
ALEPH \cite{aleph} and OPAL \cite{web} have recently
presented preliminary data on $d\sigma^L/dx$.
In Fig.~\ref{fig5}, we evaluate Eq.~(\ref{sigmal})
with our NLO FF and two-loop
$\alpha_s$ and compare the result with the ALEPH data \cite{aleph}.
The result obtained (full curve) falls short of the data by a factor of
two.
At this point, we have to keep in mind that Eq.~(\ref{sigmal}) is a LO
prediction, and one should be prepared to allow for a $K$ factor, which
is typically larger than one.
This $K$ factor may be simulated by changing the renormalization and
factorization scales.
Assuming a scale of 20~GeV, we are able to nicely describe the data.
This comparison provides us with some useful check on the gluon FF at low $x$.
A similar scale change was considered by Webber
in connection with Monte Carlo studies of this observable \cite{web}.

In Fig.~\ref{fig4}, we compared our analysis with the experimental results on
gluon fragmentation into hadrons obtained by OPAL \cite{opal} through their
three-jet analysis with the corrections of Nason and Webber \cite{opal}
applied.
A more direct comparison with the original OPAL data is feasible.
For this purpose, we calculate the ratio of the inclusive scaled energy
distribution for gluon-tagged and normal-mixture events,
$R_{\rm g.tag/n.mix}$, as a function of $x$.
This quantity is represented in our analysis by
\begin{equation}
R_{\rm g.tag/n.mix}={{D_g^h(x,Q^2)\sum_{i=1}^{2N_f}e_i^2}
\over {d\sigma/dx}},
\end{equation}
where $h$ stands for the sum over the charged hadrons and the quantity in the
denominator is defined in Eq.~(\ref{sigx}).
The result of this comparison is presented for LO (dashed line) and NLO
(solid line) in Fig.~\ref{fig6}.
We see that our model gives a good account of $R_{\rm g.tag/n.mix}$ for
medium and large $x$.

The factorization theorem guarantees that the FF characterize the
hadronization phenomena in a process-independent way, {\it i.e.}, the FF that
have been extracted from $e^+e^-$ data may also be used to make predictions for
other types of experiments, {\it e.g.}, at $\gamma\gamma$, $ep$, and hadron
colliders, fixed-target facilities, {\it etc.}
In the following, we shall make NLO predictions for inclusive photoproduction
of charged hadrons at HERA and confront them with recent high-statistics data
taken by H1 \cite{h1} and ZEUS \cite{zeus}.
We shall focus attention on the $p_T$ spectrum of the produced hadrons,
averaged over a certain $y_{\rm lab}$ interval.
We shall work at NLO in the $\overline{\rm MS}$ scheme with $N_f=5$ quark
flavours, set $\mu=M_\gamma=M_p=M_h=\xi p_T$, and adopt the photon and proton
PDF from Refs.~\cite{grv1} and \cite{cteq3}, respectively, along with our FF
for charged pions and kaons.
Unless stated otherwise, we shall choose $\xi=1$.
We shall evaluate $\alpha_s$ with $\Lambda_{\overline{\rm MS}}^{(5)}=158$~MeV
\cite{cteq3}.
As we have seen in Sect.~3, H1 and ZEUS apply different criteria to select
the photoproduction events,
which may be simulated by appropriate choices of the photon-energy cuts and
the $Q_{\rm max}^2$ value in Eq.~(\ref{wwa}).
Another difference is that the $y_{\rm lab}$ range presently covered by the H1
detector is $-1.5\le y_{\rm lab}\le1.5$, while the ZEUS detector covers
$-1.2\le y_{\rm lab}\le1.4$.

In Figs.~\ref{fig7} and \ref{fig8},
we confront the H1 and ZEUS data, respectively, with the corresponding NLO
calculations performed with our new FF sets for charged pions and kaons.
The agreement is almost perfect for $p_T\simlt4$~GeV.
For higher values of $p_T$, the central values of the measurements tend to
depart from our predictions in both directions, but, at the same time, the
experimental errors become larger.
We shall have to await more statistics in the high-$p_T$ range until we can
reach final conclusions concerning the agreement of experiment and theory.
So far, H1 and ZEUS have not detected separately charged pions and kaons.
We would like to take this opportunity and encourage these collaborations to
tackle such a separation, since this would allow us to test the factorization
theorem of QCD in greater detail.
In Fig.~\ref{fig9}, we repeat the analyses of Figs.~\ref{fig7} and \ref{fig8}
for the average
of the positively- and negatively-charged kaons (dashed lines) and compare the
outcome with the full results for H1 and ZEUS shown in
Figs.~\ref{fig7} and \ref{fig8}, respectively (solid lines).
We see that, in the case of H1 (ZEUS), the ratio of charged kaons to charged
pions ranges between 53\% and 70\% (54\% and 71\%) for $p_T$ between 1.5~GeV
and 15~GeV.
Having gained a solid amount of confidence in the validity of our
FF, we may ask the question whether the H1 and ZEUS data already show
evidence for the coexistence of the direct- and resolved-photon mechanisms.
Toward this end, we compare in Fig.~\ref{fig10} the full calculations of
Figs.~\ref{fig7} and \ref{fig8}
with the respective direct-photon contributions.
The direct-photon contribution exhibits a discontinuity at $p_T\approx3$~GeV.
This is due to the onset of the charm FF in connection with the
$\gamma g\to c\bar c$ subprocess, which is not suppressed.
In the case of H1 (ZEUS), the direct-photon contribution amounts to between
8\% and 48\% (9\% and 52\%) of the total result for $p_T$ between 1.5~GeV and
15~GeV.
Also here, we need more statistics at high $p_T$ before we can claim to have
detected the direct-photon mechanism in the inclusive photoproduction of
hadrons at HERA.
Of course, a precise experimental determination of the rapidity distribution
for perturbatively high values of $p_T$ would serve as a more direct approach
to resolve this issue.
The H1 Collaboration has already undertaken a first step in that direction
\cite{h1}.
Another interesting issue is whether the gluon content of the resolved photon
has been established by the data.
In the case of H1 (ZEUS), it makes up between 67\% and 10\% (66\% and 7\%)
of the resolved contribution and between 61\% and 5\% (61\% and 3\%)
of the total one for $p_T$ between 1.5~GeV and 15~GeV.
Thus, it seems that, in the absence of the gluon density inside the photon,
the theoretical prediction would significantly fall short of the data at low
$p_T$.
Again, the shape of the measured rapidity spectrum would serve as an additional
discriminator.

Finally, we investigate if our FF satisfy the momentum sum rules.
Guided by the idea that a given outgoing parton, $a$, will fragment with
100\% likelihood into some hadron, $h$, and that momentum is conserved during
the fragmentation process, we expect that
\begin{equation}
\label{sumrule}
\sum_h\int_0^1dx\,xD_a^h(x,Q^2)=1
\end{equation}
holds for any value of $Q^2$.
In our analysis, the left-hand-side of Eq.~(\ref{sumrule}) should be smaller
than unity, since we consider only pions, kaons, protons, and antiprotons,
which does not exhaust the spectrum of all possible hadrons.
We approximate the contributions due to neutral pions, neutral kaons, and
protons/antiprotons by
\begin{eqnarray}
\label{pinull}
D_a^{\pi^0}(x,Q^2)&=&{1\over2}D_a^{\pi^++\pi^-}(x,Q^2),\\
\label{knull}
D_a^{K^0+\bar K^0}(x,Q^2)&=&D_a^{K^++K^-}(x,Q^2),\\
\label{ppbar}
D_a^{p+\bar p}(x,Q^2)&=&f(x)D_a^{\pi^++\pi^-}(x,Q^2),
\end{eqnarray}
respectively, where $f(x)$ was introduced in Eq.~(\ref{fx}).
Equation~(\ref{knull}) is in good agreement with data at various
energies \cite{argus,k0},
while Eq.~(\ref{pinull}) follows on from SU(2) symmetry.
We take the lower limit of integration in Eq.~(\ref{sumrule}) to be 0.02.
This is presumably too small for the low energies $\sqrt s=2,4,10$~GeV,
which might explain why the sum-rule values are in some cases larger than one
at these energies, in particular for our LO fit.
In Table~II, we list the results obtained with our LO and NLO FF
for $Q=\sqrt2$, 5, 10, 91, and 200~GeV.
The accuracy of our results is limited due to the uncertainty in the
assumptions~(\ref{pinull})--(\ref{ppbar}), and we estimate the error to be of
the order of 10\%.

\begin{table} {TABLE~II. Left-hand side of Eq.~(\ref{sumrule}) at NLO for
$Q=\sqrt 2$, 5, 10, 91, and 200~GeV.
The numbers in parentheses are evaluated with our LO set.
We integrate over $0.02<x<0.98$ and sum over
pions, kaons, protons, and antiprotons.}\\[1ex]
\begin{tabular}{|c|c|c|c|c|c|} \hline
$a$ & \multicolumn{5}{c|}{$Q$ [GeV]} \\
\cline{2-6} & $\sqrt2$ & 4 & 10 & 91 & 200 \\
\hline
$u$ & 0.83 (1.10) & 0.86 (1.03) & 0.86 (0.98) & 0.81 (0.89) & 0.79 (0.86) \\
$d$ & 0.95 (1.08) & 0.96 (1.02) & 0.94 (0.97) & 0.88 (0.88) & 0.86 (0.85) \\
$s$ & 1.08 (1.70) & 1.05 (1.52) & 1.02 (1.42) & 0.94 (1.23) & 0.91 (1.18) \\
$c$ & --          & 0.94 (1.14) & 0.92 (1.08) & 0.86 (0.96) & 0.83 (0.92) \\
$b$ & --          & --          & 0.92 (1.11) & 0.85 (0.98) & 0.83 (0.94) \\
$g$ & 0.96 (0.93) & 0.96 (0.87) & 0.91 (0.82) & 0.79 (0.70) & 0.75 (0.67) \\
\hline
\end{tabular}
\end{table}

\section{Summary and Conclusions}

We presented new sets of FF for charged pions and kaons, both at LO and NLO.
They were fitted to data on inclusive charged-pion and -kaon production in
$e^+e^-$ annihilation taken by TPC \cite{tpc} at PEP ($\sqrt s=29$~GeV) and by
ALEPH \cite{aleph,aleph1} at LEP.
In their charged-hadron sample \cite{aleph}, ALEPH discriminated between
events with charm, bottom, and light-flavour fragmentation.
This enabled us to treat all partons independently and to properly incorporate
the charm and bottom thresholds.
We just imposed the conditions $D_d^{\pi^++\pi^-}=D_u^{\pi^++\pi^-}$ and
$D_s^{K^++K^-}=D_u^{K^++K^-}$, which are quite natural, since we treated the
$u$, $d$, and $s$ quarks as massless, so that they equally participate in the
strong interaction.
This constitutes a considerable improvement of our previous analysis
\cite{bkk} based just on the TPC data, where we could only distinguish between
valence-quark, sea-quark, and gluon fragmentation and neglected the
heavy-flavour thresholds.
In order to control the gluon FF, we exploited recent data on inclusive
charged-hadron production in tagged three-jet events by OPAL \cite{opal} and
similar data for longitudinal electron polarization by ALEPH \cite{aleph}.
Due to the sizeable energy gap between PEP and LEP, we were also sensitive to
the scaling violation in the fragmentation process.
This allowed us to extract a value for $\Lambda_{\overline{\rm MS}}^{(5)}$.
We found $\Lambda_{\overline{\rm MS}}^{(5)}=195$~MeV ($107$~MeV) in NLO (LO),
in good agreement with the global analysis of LEP data \cite{heb}.

Although our FF were only fitted to the TPC \cite{tpc} and ALEPH data
\cite{aleph,aleph1},
it turned out that they lead to an excellent description of a wealth of other
$e^+e^-$ data on inclusive charged-hadron production,
ranging from $\sqrt s=5.2$~GeV to LEP energy
\cite{dasp,argus,markii,cello,tasso,amy,delphi}.
In fact, we always obtained $\chi^2_{\rm d.o.f.}$ values of order unity.
The only exception, with $\chi^2_{\rm d.o.f.}\approx9$, occurred in the
longitudinal cross section \cite{aleph}, which is of $O(\alpha_s)$ and suffers
from a considerable scale ambiguity.

In order to test quantitatively the factorization theorem of fragmentation in
the QCD-improved parton model at the quantum level,
we made NLO predictions for the $p_T$ spectrum of charged hadrons produced
inclusively in the scattering of quasi-real photons on protons under HERA
conditions, and confronted them with new high-statistics data by H1 \cite{h1}
and ZEUS \cite{zeus}.
Also here, the outcome of the comparison was very encouraging, especially
in the lower $p_T$ range, where the experimental errors are smallest.
We found first evidence for the interplay between the direct- and
resolved-photon mechanisms in the inclusive photoproduction of hadrons.
Our comparisons also support the notion that, during a fraction of time, the
photon can interact with the partons inside the proton like a gluon.

\bigskip
\centerline{\bf ACKNOWLEDGMENTS}
\smallskip\noindent
We thank Glen Cowan and Cristobal Padilla for detailed
information on Ref.~\cite{aleph}.
Furthermore, we thank Michal Kasprzak for making available to us the latest
ZEUS data on inclusive photoproduction of charged hadrons prior to their
official publication \cite{zeus}.
One of us (GK) thanks the Theory Group of the Werner-Heisenberg-Institut for
the hospitality extended to him during a visit when this paper was finalized.

\begin{appendix}

\section{Parameterizations}

For the reader's convenience, we shall present here simple parameterizations
of the $x$ and $Q^2$ dependence of our FF.\footnote{%
A FORTRAN subroutine that returns the FF for given $x$ and $Q^2$ may be
obtained from the authors via e-mail (binnewie@ips107.desy.de,
kniehl@vms.mppmu.mpg.de).}
As usual, we introduce the scaling variable
\begin{equation}
\label{sbar}
\bar{s}=\ln{\frac{\ln{(Q^2/\Lambda^2)}}{\ln{(Q^2_0/\Lambda^2)}}}.
\end{equation}
For $\Lambda$ we use the $\overline{\rm MS}$ value appropriate to $N_f=5$
flavours, since the parameterization would not benefit from the incorporation
of discontinuities in $\bar{s}$.
$\Lambda_{\overline{\rm MS}}^{(5)}$ is determined from our fit to be
107~MeV (195~MeV) in LO (NLO).
Similarly to Eqs.~(\ref{pilo})--(\ref{kanlo}), we use three different values
for $Q_0$, namely
\begin{eqnarray}
Q_0&=&\left\{\begin{array}{l@{\quad\mbox{if}\quad}l}
\sqrt2~\mbox{GeV}, & a=u,d,s,g\\
m(\eta_c)=2.9788~\mbox{GeV}, & a=c\\
m(\Upsilon)=9.46037~\mbox{GeV}, & a=b
\end{array}\right.\;.
\end{eqnarray}
This leads to three different definitions of $\bar s$.
For definiteness, we use the symbol $\bar s_c$ for charm and $\bar s_b$ for
bottom along with $\bar s$ for the residual partons.

We parameterize our FF by simple functions in $x$ with coefficients
which we write as polynomials in $\bar s$, $\bar s_c$, and $\bar s_b$.
We find that the template
\begin{equation}
\label{temp}
D(x,Q^2)=Nx^{\alpha}(1-x)^{\beta}
\end{equation}
is sufficiently flexible,
except for $D_b^{(\pi,LO)}$, $D_b^{(\pi,NLO)}$, and $D_g^{(K,NLO)}$,
where we include an additional factor $(1+\gamma/x)$ on the right-hand side of
Eq.~(\ref{temp}).
For $\bar s=\bar s_c=\bar s_b=0$, the parameterizations agree with
the respective ans\"atze in Eqs.~(\ref{pilo})--(\ref{kanlo}).
The charm and bottom parameterizations must be put to zero by hand for
$\bar s_c<0$ and $\bar s_b<0$, respectively.

We list below the parameters to be inserted in Eq.~(\ref{temp})
for charged pions and kaons both at LO and NLO.
The resulting parameterizations correctly describe the AP evolution up to 10\%
for $Q_0\le Q\le150$~GeV and $x\ge0.1$.
Deviations in excess of 10\% may occur for $x<0.1$.
\begin{enumerate}
\item LO FF for $(\pi^++\pi^-)$:
\begin{itemize}
\item $D_u^{(\pi,LO)}(x,Q^2)=D_d^{(\pi,LO)}(x,Q^2)$:
\begin{eqnarray}
N&\n=\n&1.090-1.085\bar s +0.700\bar s^2 -0.217\bar s^3\nonumber\\
\alpha&\n=\n& -0.842 -0.920\bar s +0.532\bar s^2 -0.148\bar s^3\nonumber\\
\beta&\n=\n&1.430 +0.434\bar s +0.035\bar s^2 +0.060\bar s^3
\end{eqnarray}
\item $D_s^{(\pi,LO)}(x,Q^2)$:
\begin{eqnarray}
N&\n=\n&3.470-2.340\bar s +0.309\bar s^2 +0.120\bar s^3\nonumber\\
\alpha&\n=\n&-1.037 -0.301\bar s -0.038\bar s^2 +0.031\bar s^3\nonumber\\
\beta&\n=\n&3.880 +0.866\bar s -0.137\bar s^2 +0.058\bar s^3
\end{eqnarray}
\item $D_c^{(\pi,LO)}(x,Q^2)$:
\begin{eqnarray}
N&\n=\n&4.520-3.966\bar s_c +1.453\bar s_c^2 -0.261\bar s_c^3\nonumber\\
\alpha&\n=\n& -0.898 -0.382\bar s_c +0.027\bar s_c^3\nonumber\\
\beta&\n=\n&4.690 +0.742\bar s_c
\end{eqnarray}
\item $D_b^{(\pi,LO)}(x,Q^2)$:
\begin{eqnarray}
N&\n=\n&3.620-9.669\bar s_b+13.160\bar s_b^2-5.406\bar s_b^3\nonumber\\
\alpha&\n=\n&-1.128 -1.613\bar s_b +2.715\bar s_b^3\nonumber\\
\beta&\n=\n&7.120 -1.081\bar s_b +0.936\bar s_b^2 +1.678\bar s_b^3\nonumber\\
\gamma&\n=\n& -0.164\bar s_b +0.208\bar s_b^2
\end{eqnarray}
\item $D_g^{(\pi,LO)}(x,Q^2)$:
\begin{eqnarray}
N&\n=\n&6.200-13.100\bar s+10.940\bar s^2-3.395\bar s^3\nonumber\\
\alpha&\n=\n&-0.351-1.542\bar s +0.319\bar s^2 +0.074\bar s^3\nonumber\\
\beta&\n=\n&3.000+1.567\bar s -0.725\bar s^2 +0.465\bar s^3
\end{eqnarray}
\end{itemize}
\item NLO FF for $(\pi^++\pi^-)$:
\begin{itemize}
\item $D_u^{(\pi,NLO)}(x,Q^2)=D_d^{(\pi,NLO)}(x,Q^2)$:
\begin{eqnarray}
N&\n=\n&1.160-1.292\bar s +0.975\bar s^2 -0.341\bar s^3\nonumber\\
\alpha&\n=\n&-0.762-1.392\bar s+1.151\bar s^2 -0.429\bar s^3\nonumber\\
\beta&\n=\n&1.470 +0.752\bar s -0.350\bar s^2 +0.175\bar s^3
\end{eqnarray}
\item $D_s^{(\pi,NLO)}(x,Q^2)$:
\begin{eqnarray}
N&\n=\n&3.960-3.148\bar s+1.198\bar s^2 -0.267\bar s^3\nonumber\\
\alpha&\n=\n&-0.825 -0.552\bar s +0.206\bar s^2 -0.070\bar s^3\nonumber\\
\beta&\n=\n&4.380 +0.940\bar s -0.245\bar s^2 +0.082\bar s^3
\end{eqnarray}
\item $D_c^{(\pi,NLO)}(x,Q^2)$:
\begin{eqnarray}
N&\n=\n&3.690 -3.285\bar s_c +1.383\bar s_c^2 -0.305\bar s_c^3\nonumber\\
\alpha&\n=\n& -0.873 -0.481\bar s_c +0.085\bar s_c^2 \nonumber\\
\beta&\n=\n&5.060 +0.685\bar s_c -0.032\bar s_c^2 +0.038\bar s_c^3
\end{eqnarray}
\item $D_b^{(\pi,NLO)}(x,Q^2)$:
\begin{eqnarray}
N&\n=\n&4.000-12.965\bar s_b+18.913\bar s_b^2-10.853\bar s_b^3\nonumber\\
\alpha&\n=\n&-1.037 -1.705\bar s_b -0.865\bar s_b^2 +1.423\bar s_b^3\nonumber\\
\beta&\n=\n&7.750 -2.306\bar s_b+1.877\bar s_b^2 +0.044\bar s_b^3\nonumber\\
\gamma&\n=\n& -0.116\bar s_b
\end{eqnarray}
\item $D_g^{(\pi,NLO)}(x,Q^2)$:
\begin{eqnarray}
N&\n=\n&5.110-9.419\bar s+7.144\bar s^2-2.069\bar s^3\nonumber\\
\alpha&\n=\n&-0.287-2.011\bar s+1.211\bar s^2 -0.377\bar s^3\nonumber\\
\beta&\n=\n&2.610+2.336\bar s-1.580\bar s^2 +0.756\bar s^3
\end{eqnarray}
\end{itemize}
\item LO FF for $(K^++K^-)$:
\begin{itemize}
\item $D_u^{(K,LO)}(x,Q^2)=D_s^{(K,LO)}(x,Q^2)$:
\begin{eqnarray}
N&\n=\n&0.380 -0.123\bar s -0.015\bar s^2 +0.010\bar s^3\nonumber\\
\alpha&\n=\n&-1.234 -0.106\bar s -0.035\bar s^2 +0.011\bar s^3\nonumber\\
\beta&\n=\n&1.040 +0.708\bar s -0.086\bar s^2 +0.048\bar s^3
\end{eqnarray}
\item $D_d^{(K,LO)}(x,Q^2)$:
\begin{eqnarray}
N&\n=\n&1.110 -0.674\bar s +0.095\bar s^2 +0.028\bar s^3\nonumber\\
\alpha&\n=\n& -0.942 -0.214\bar s -0.067\bar s^2 +0.041\bar s^3\nonumber\\
\beta&\n=\n&2.810 +0.852\bar s -0.172\bar s^2 +0.090\bar s^3
\end{eqnarray}
\item $D_c^{(K,LO)}(x,Q^2)$:
\begin{eqnarray}
N&\n=\n&0.550 -0.392\bar s_c +0.088\bar s_c^2\nonumber\\
\alpha&\n=\n&-0.764 -0.378\bar s_c\nonumber\\
\beta&\n=\n&2.690 +0.713\bar s_c-0.103\bar s_c^2+0.064\bar s_c^3
\end{eqnarray}
\item $D_b^{(K,LO)}(x,Q^2)$:
\begin{eqnarray}
N&\n=\n&0.720 -0.495\bar s_b+0.205\bar s_b^3\nonumber\\
\alpha&\n=\n&-0.792 -0.296\bar s_b -0.261\bar s_b^2 +0.421\bar s_b^3\nonumber\\
\beta&\n=\n&2.810+0.686\bar s_b
\end{eqnarray}
\item $D_g^{(K,LO)}(x,Q^2)$:
\begin{eqnarray}
N&\n=\n&0.430 -0.915\bar s +0.885\bar s^2 -0.316\bar s^3\nonumber\\
\alpha&\n=\n&-0.374-2.368\bar s+1.475\bar s^2 -0.317\bar s^3\nonumber\\
\beta&\n=\n&2.690+1.137\bar s -0.934\bar s^2 +0.698\bar s^3
\end{eqnarray}
\end{itemize}
\item NLO FF for $(K^++K^-)$:
\begin{itemize}
\item $D_u^{(K,NLO)}(x,Q^2)=D_s^{(K,NLO)}(x,Q^2)$:
\begin{eqnarray}
N&\n=\n&0.300 +0.039\bar s -0.189\bar s^2 +0.070\bar s^3\nonumber\\
\alpha&\n=\n&-0.978 -0.304\bar s -0.004\bar s^2 -0.004\bar s^3\nonumber\\
\beta&\n=\n&1.010+1.193\bar s -0.654\bar s^2 +0.223\bar s^3
\end{eqnarray}
\item $D_d^{(K,NLO)}(x,Q^2)$:
\begin{eqnarray}
N&\n=\n&1.050 -0.546\bar s +0.071\bar s^2 +0.009\bar s^3\nonumber\\
\alpha&\n=\n&-0.804 -0.347\bar s +0.023\bar s^2 +0.007\bar s^3\nonumber\\
\beta&\n=\n&2.540 +0.919\bar s -0.321\bar s^2 +0.138\bar s^3
\end{eqnarray}
\item $D_c^{(K,NLO)}(x,Q^2)$:
\begin{eqnarray}
N&\n=\n&0.680 -0.415\bar s_c +0.092\bar s_c^2\nonumber\\
\alpha&\n=\n&-0.753 -0.484\bar s_c +0.099\bar s_c^2\nonumber\\
\beta&\n=\n&2.590 +0.661\bar s_c
\end{eqnarray}
\item $D_b^{(K,NLO)}(x,Q^2)$:
\begin{eqnarray}
N&\n=\n&0.580 -0.381\bar s_b +0.114\bar s_b^3\nonumber\\
\alpha&\n=\n&-0.843 -0.414\bar s_b-0.120\bar s_b^2 +0.242\bar s_b^3\nonumber\\
\beta&\n=\n&2.880 +0.614\bar s_b
\end{eqnarray}
\item $D_g^{(K,NLO)}(x,Q^2)$:
\begin{eqnarray}
N&\n=\n&0.330 -0.196\bar s -0.056\bar s^2 +0.057\bar s^3\nonumber\\
\alpha&\n=\n&-0.351 -0.026\bar s -0.480\bar s^2 +0.148\bar s^3\nonumber\\
\beta&\n=\n&0.650+2.326\bar s -0.840\bar s^2 +0.314\bar s^3\nonumber\\
\gamma&\n=\n&1.033\bar s -0.479\bar s^2
\end{eqnarray}
\end{itemize}
\end{enumerate}

\end{appendix}

\newpage

\vskip-6cm

\begin{figure}

\centerline{\bf FIGURE CAPTIONS}

\caption{\protect\label{fig1} $x$ dependence of the FF at $Q^2=100$~GeV$^2$
for (a) charged pions at NLO, (b) charged pions at LO,
(c) charged kaons at NLO, and (d) charged kaons at LO.\hskip5cm}

\vskip-.2cm

\caption{\protect\label{fig2} Differential cross section of inclusive
charged-pion production at LO (dashed lines) and NLO (solid lines) as a
function of $x$ at $\protect\sqrt{s}=5.2$ \protect\cite{dasp},
9.98 \protect\cite{argus}, 29 \protect\cite{tpc}, 34 \protect\cite{tasso}, and
91.2~GeV \protect\cite{aleph1}.
Upper curves correspond to lower energies.
The theoretical calculations are compared with the respective experimental
data.\hskip5cm}

\vskip-.2cm

\caption{\protect\label{fig3} Same as in Fig.~\protect\ref{fig2} for
charged kaons.\hskip8cm}

\vskip-.2cm

\caption{\protect\label{fig4} Differential cross section of inclusive
charged-hadron production at LO (dashed lines) and NLO (solid lines) as a
function of $x$ at $\protect\sqrt{s}=29$ \protect\cite{markii},
35 \protect\cite{cello}, 55.2 \protect\cite{amy}, and
91.2~GeV \protect\cite{aleph,opal}.
Upper curves correspond to lower energies.
The downmost four curves all refer to LEP energy.
The second and third of them correspond to the contributions due to
the $u,d,s$ quarks and the $b$ quark \protect\cite{aleph},
respectively. The fourth spectrum shows the $x$ dependence of the gluon FF
as measured by OPAL \protect\cite{opal}.
The theoretical calculations are compared with the respective experimental
data.\hskip5cm}

\vskip-.2cm

\caption{\protect\label{fig5} Differential longitudinal cross section of
inclusive charged-hadron production at NLO as a function of $x$ at
$\protect\sqrt{s}=91.2$~GeV.
The dashed line is obtained by choosing the renormalization and fragmentation
scales to be 20~GeV.
The theoretical calculations are compared with the experimental
data \protect\cite{aleph}.\hskip5cm}

\vskip-.2cm

\caption{\protect\label{fig6} Ratio of the inclusive scaled energy
distribution for gluon-tagged and normal-mixture events at LO (dashed line)
and NLO (solid line) as a function of $x$ for $\protect\sqrt{s}=91.2$~GeV.
The theoretical calculations are compared with the experimental
data \protect\cite{opal}.\hskip5cm}

\vskip-.2cm

\caption{\protect\label{fig7} The $p_T$ spectrum of inclusive charged-hadron
production as measured by ZEUS \protect\cite{zeus} is compared with the NLO
calculation in the $\overline{\rm MS}$ scheme with $N_f=5$ flavours using the
photon and proton PDF of Refs.~\protect\cite{grv1} and \protect\cite{cteq3},
respectively.
The dashed/solid/dotted curves correspond to the choices $\xi=0.5/1/2$.
\hskip5cm}

\vskip-.2cm

\caption{\protect\label{fig8} Same as in Fig.~\protect\ref{fig7} for
H1 \protect\cite{h1}.
\hskip10cm}

\vskip-.2cm

\caption{\protect\label{fig9} Comparison of the charged-kaon contributions
(dashed lines) with the full results (solid lines) for the theoretical
calculations in Figs.~\protect\ref{fig7} and \protect\ref{fig8}.
For better separation, the
charged-kaon contributions are divided by a factor of two.
The upper (lower) curves correspond to H1 (ZEUS) conditions.
\hskip5cm}

\vskip-.2cm

\caption{\protect\label{fig10} Comparison of the direct-photon contributions
(dashed lines) with the full results (solid lines) for the theoretical
calculations in Figs.~\protect\ref{fig7} and \protect\ref{fig8}.
The steps in the direct-photon contributions are due to the opening
of the $\gamma g\to c\bar c$ subprocess at the charm-quark threshold.
The upper (lower) curves correspond to H1 (ZEUS) conditions.
\hskip5cm}

\end{figure}

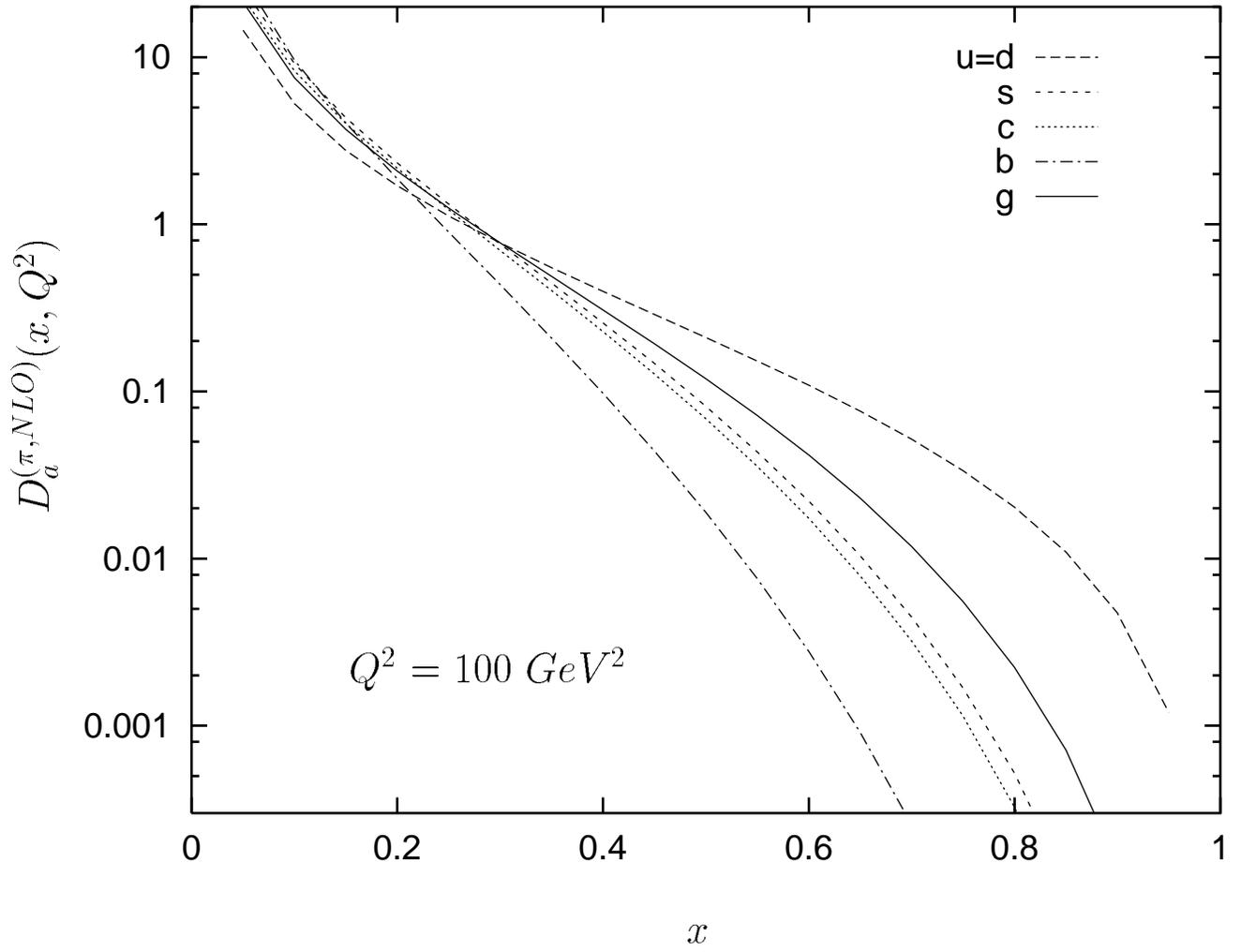
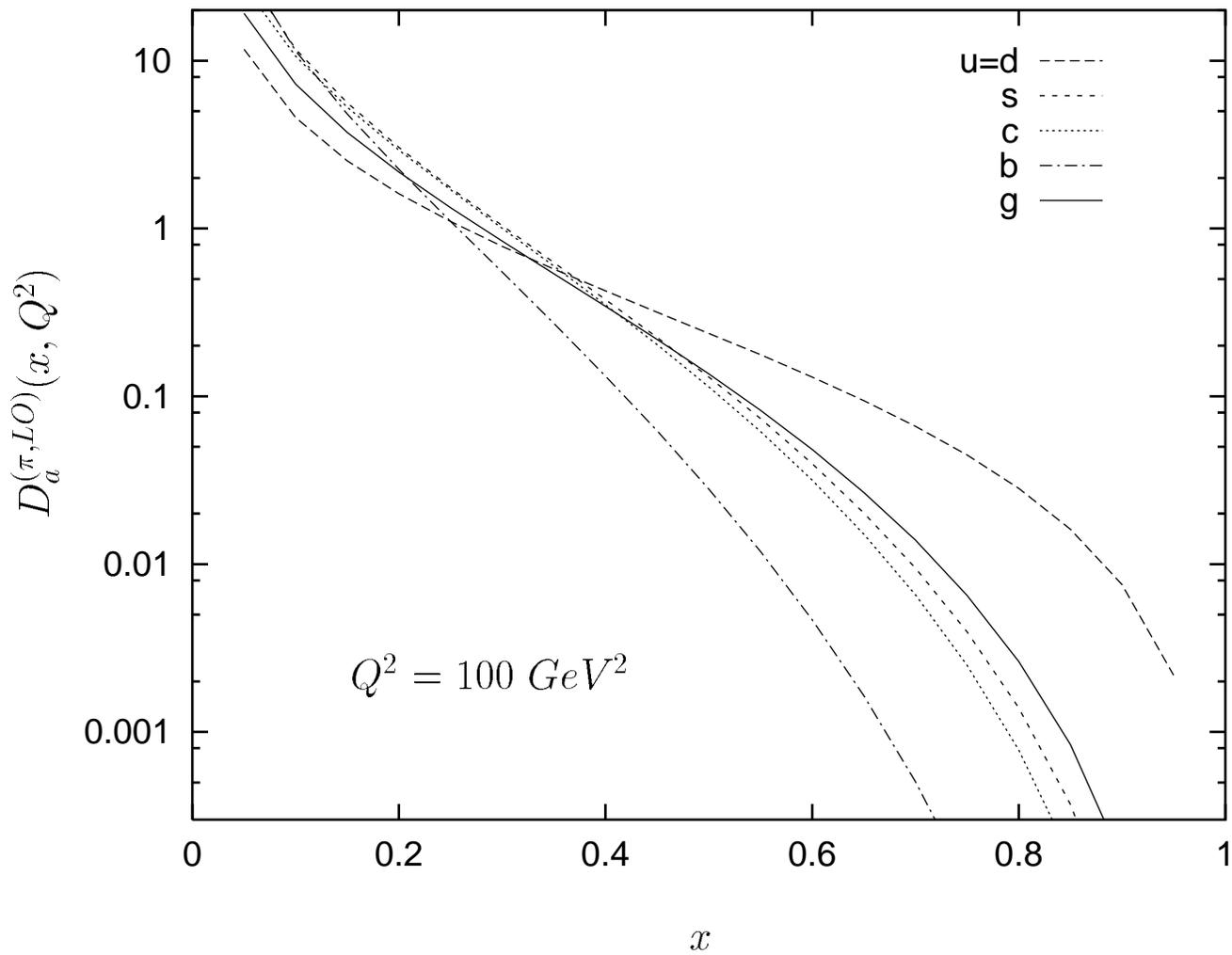
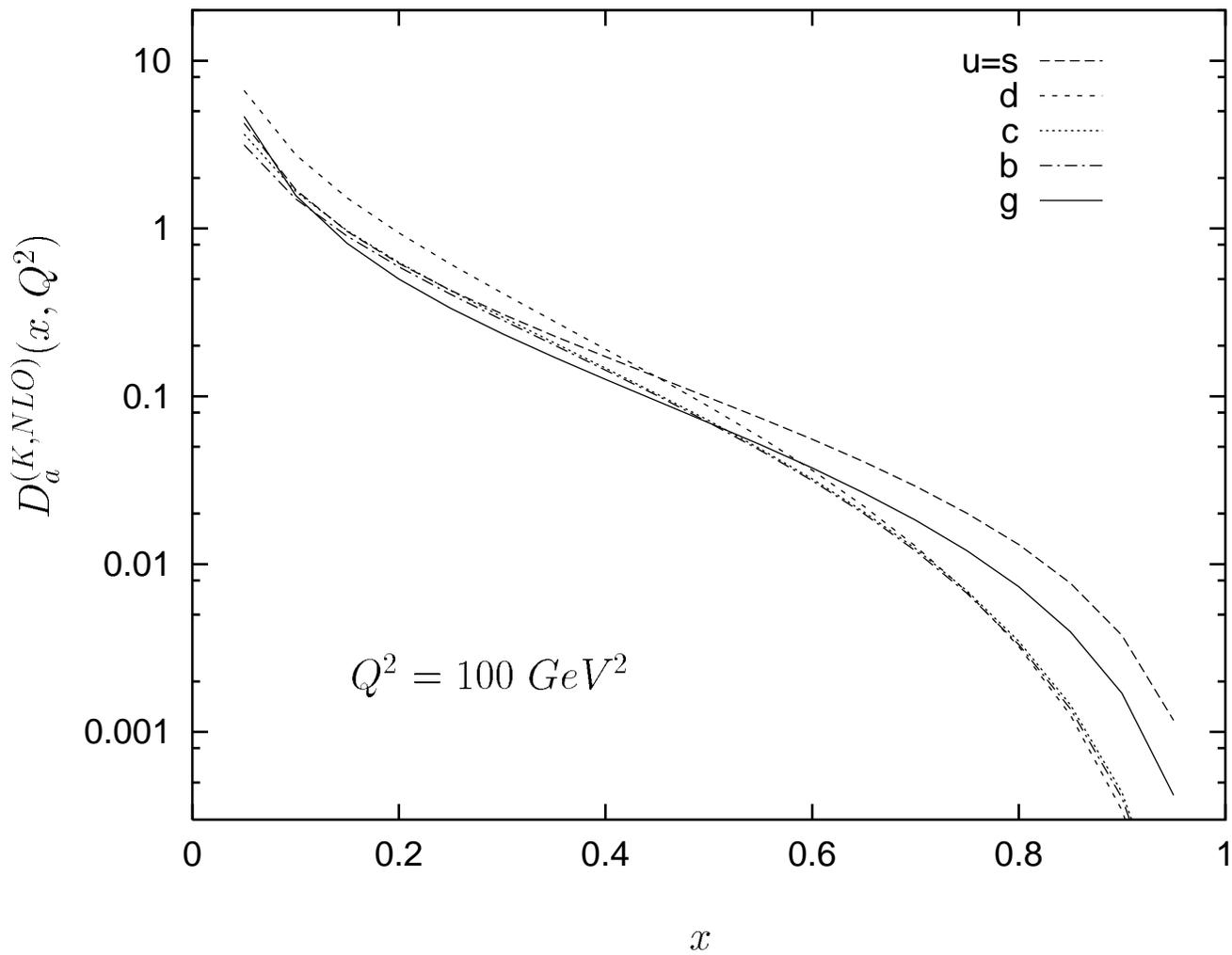
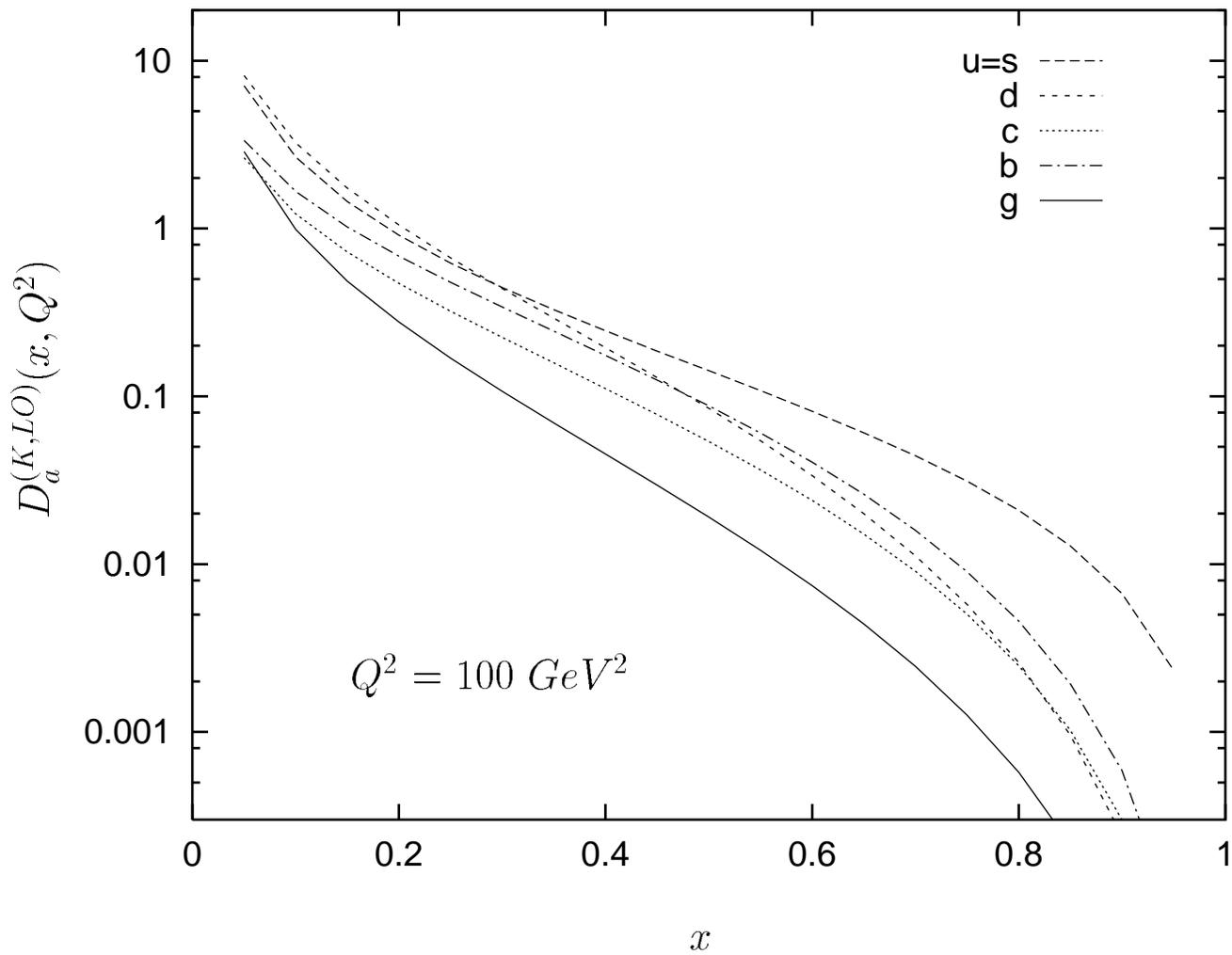
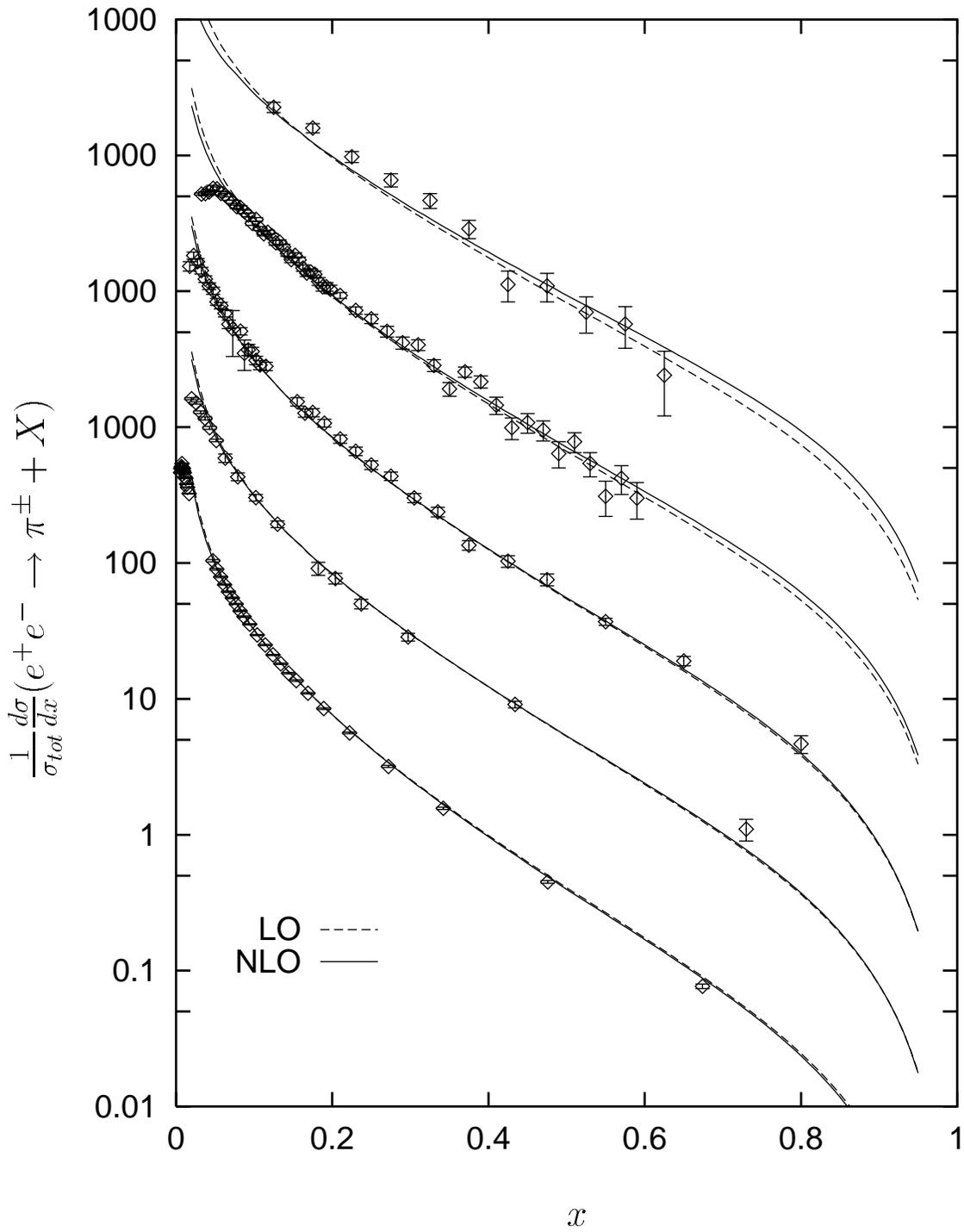
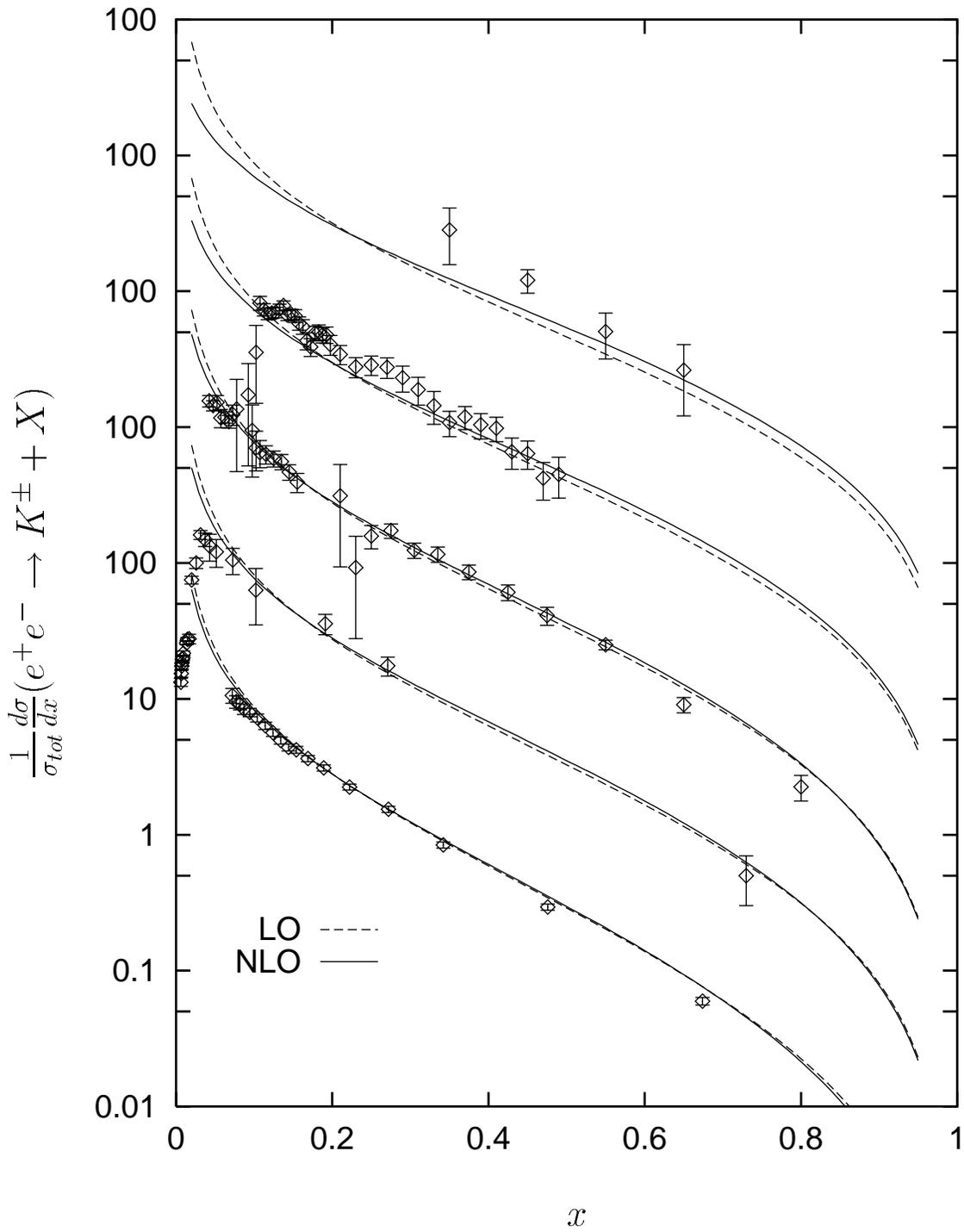
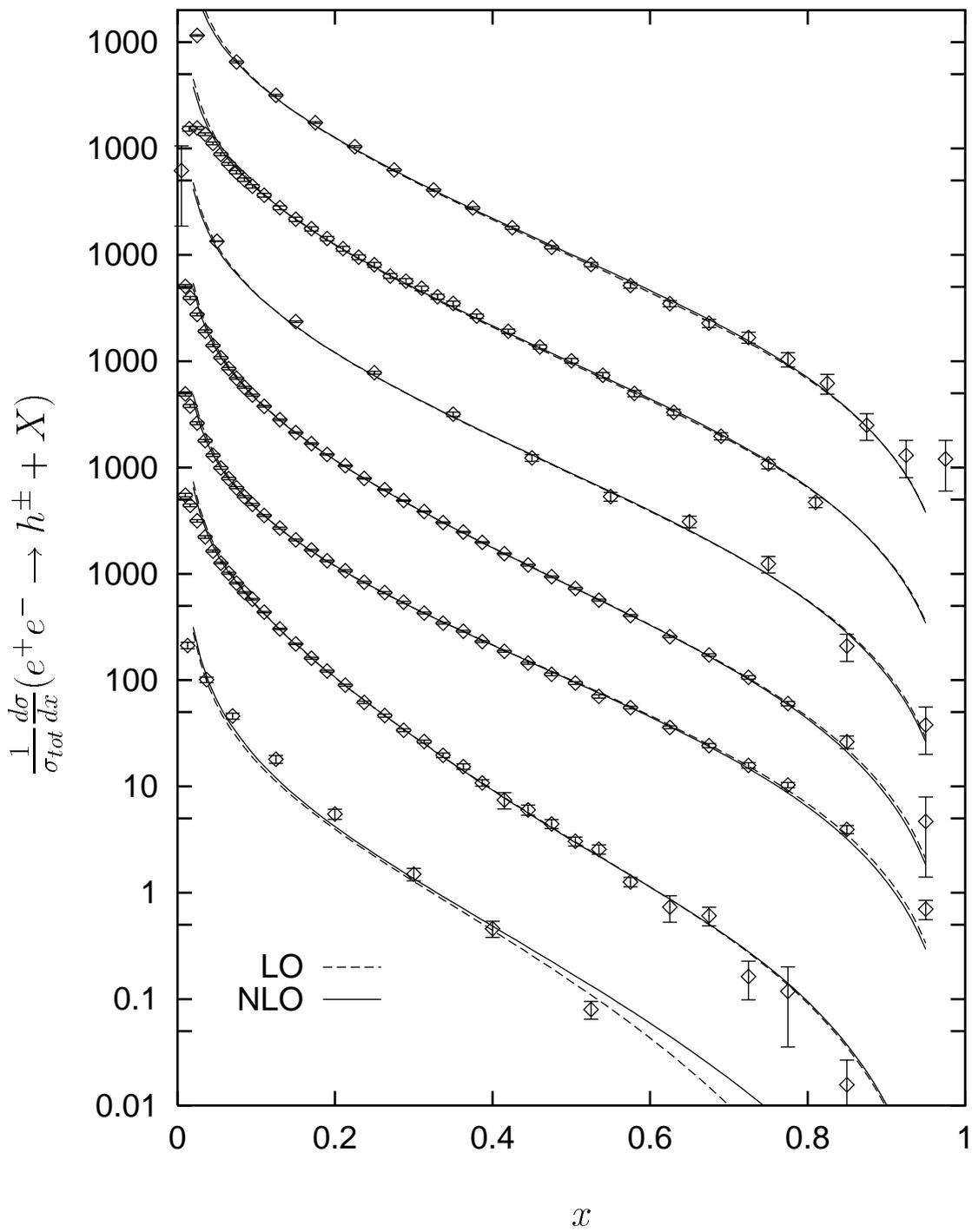
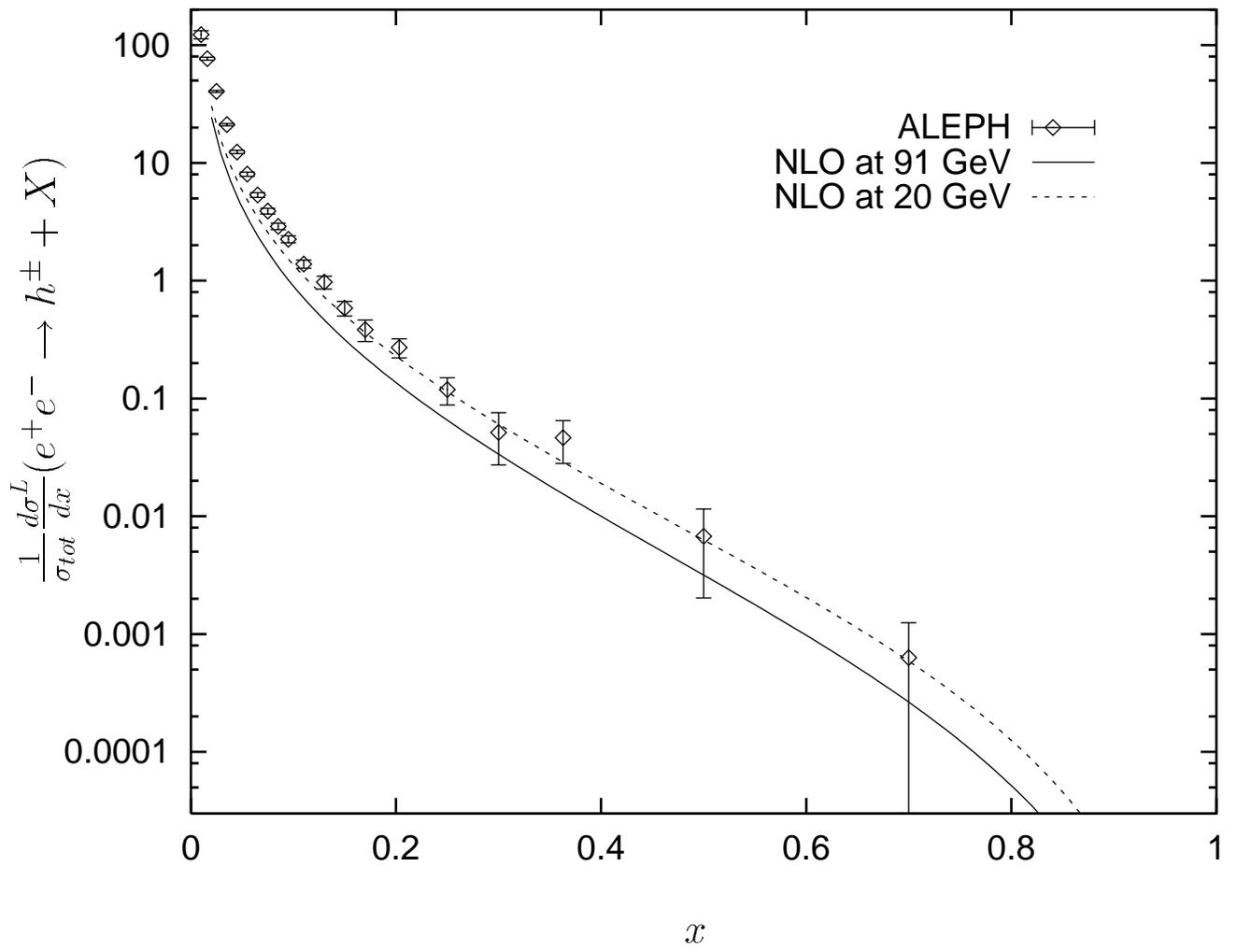
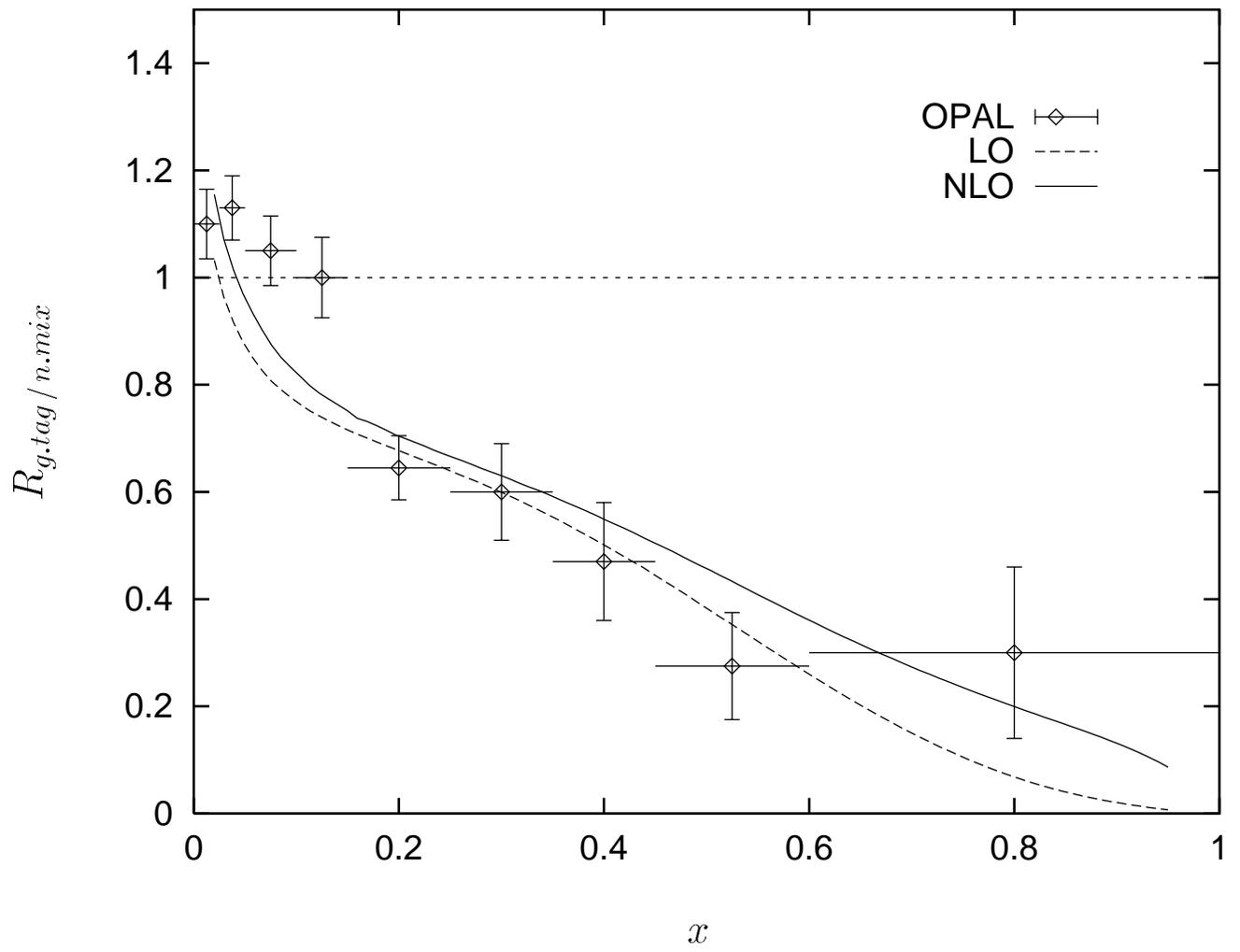
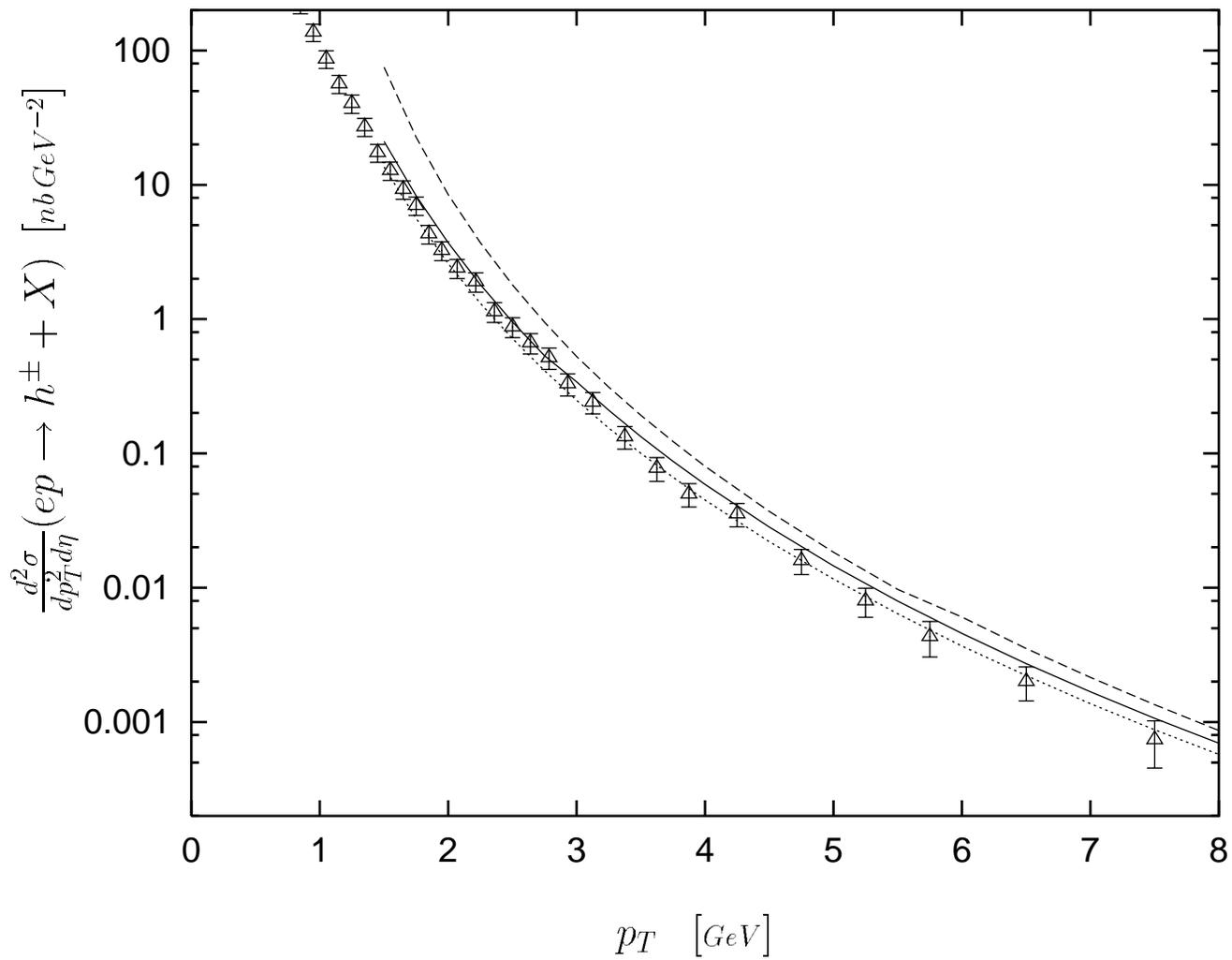
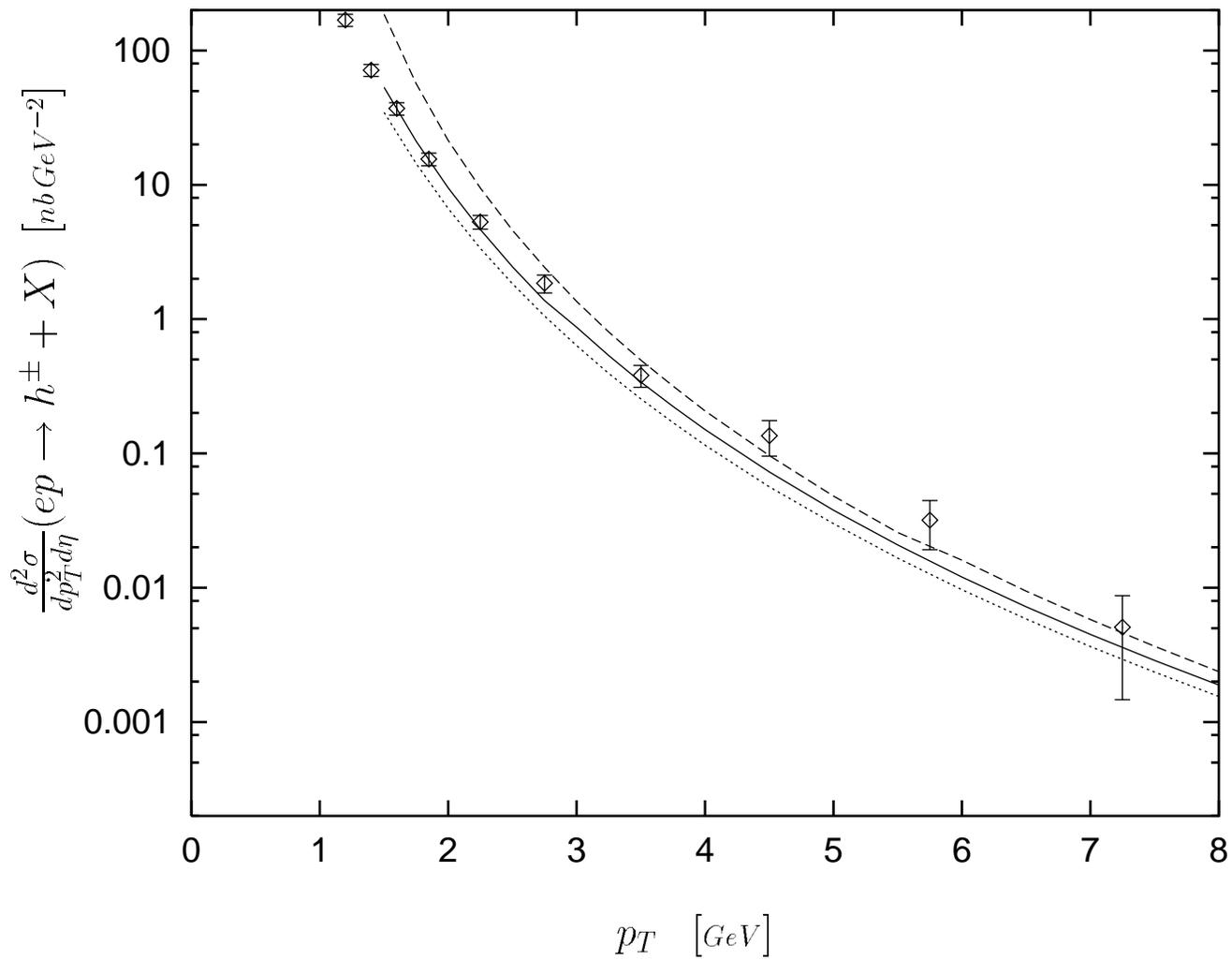
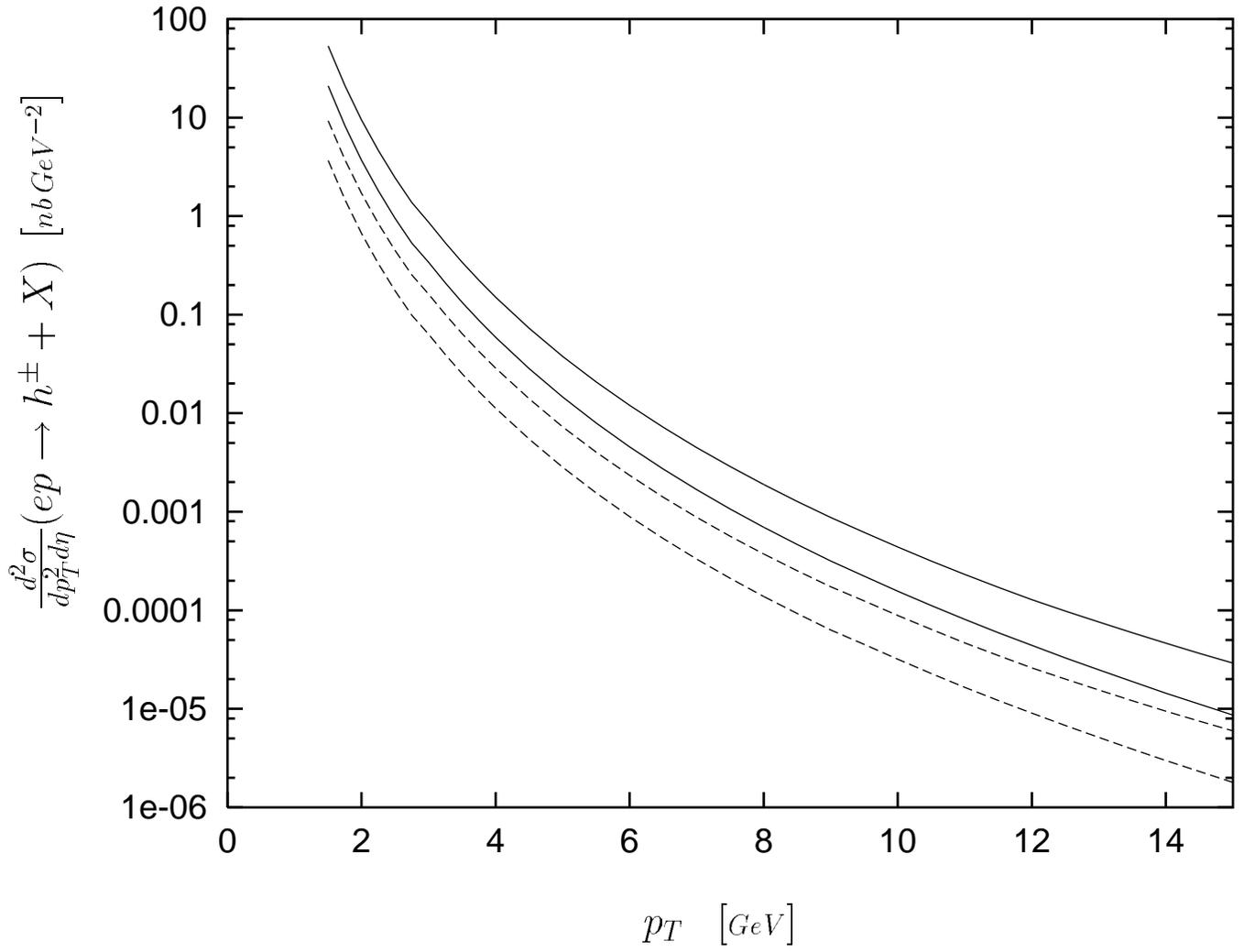
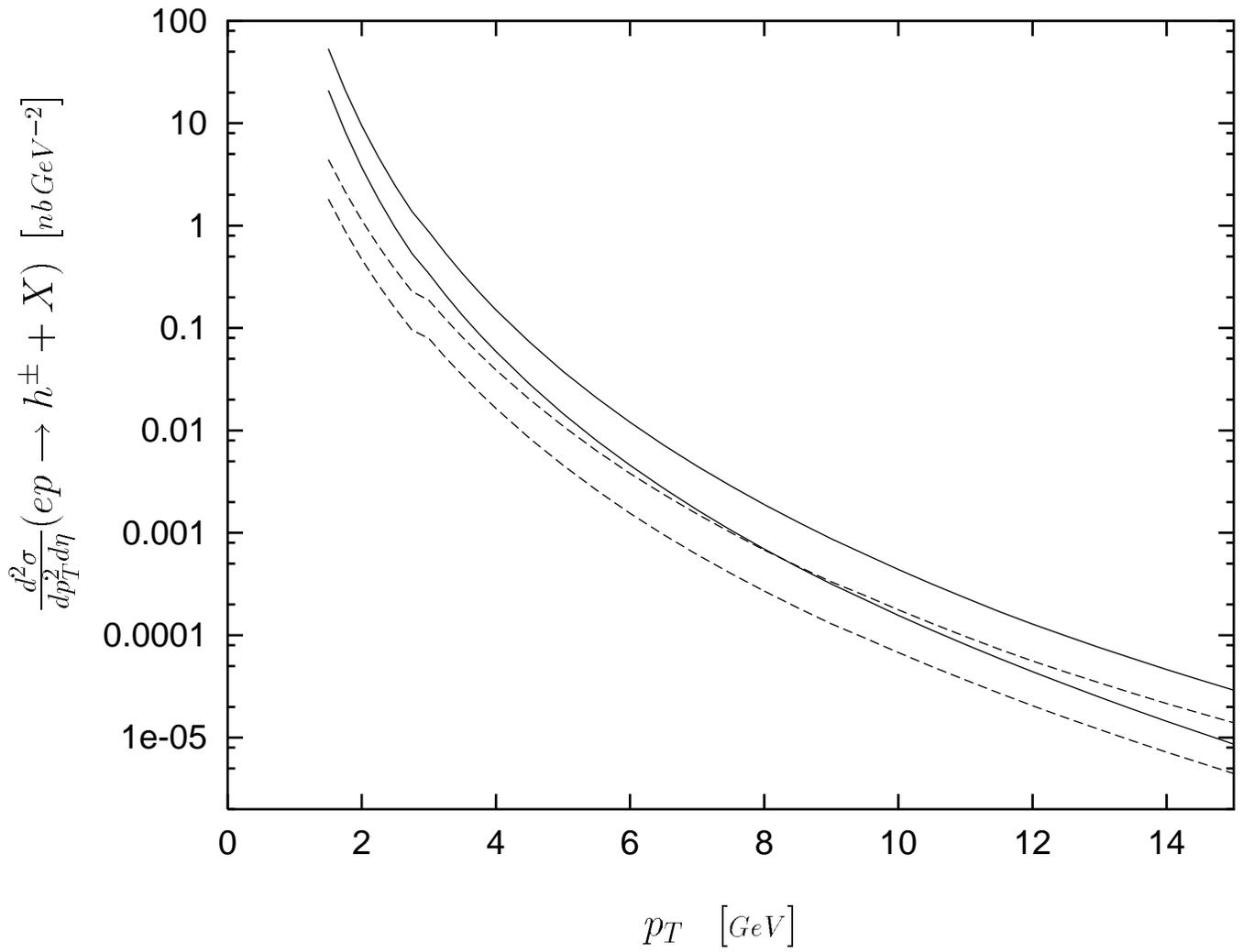
\end{document}